%% file: main.tex
\documentclass[fleqn,usenatbib]{mnras}

\usepackage{newtxtext,newtxmath}

\usepackage[T1]{fontenc}

\usepackage{graphicx}	
\usepackage{amsmath}	
\usepackage{CJKutf8}    
\usepackage{orcidlink}  
\usepackage{lastpage}  

\newcommand{\petar}{\textsc{petar}}

\newcommand{\dbscan}{\textsc{dbscan}}
\newcommand{\hdbscan}{\textsc{hdbscan}}
\newcommand{\stargo}{\textsc{stargo}}

\newcommand{\flag}[1]{\texttt{\lowercase{#1}}}
\newcommand{\galpy}{\flag{galpy}}
\newcommand{\galaxia}{\flag{galaxia}}

\title[IMF-aware Membership Selection Functions]{IMF-aware $N$-body forward modelling of membership selection functions for young open clusters}

\author[Y.-Y. Shen et al.]{Yue-Yue Shen (\begin{CJK*}{UTF8}{gbsn}沈越月\ignorespacesafterend\end{CJK*}) \orcidlink{0000-0001-8636-2990}, $^{1,2,3}$
Chao Liu (\begin{CJK*}{UTF8}{gbsn}刘超\ignorespacesafterend\end{CJK*}) \orcidlink{0000-0002-1802-6917}, $^{1,2,4,5}$\thanks{E-mail: liuchao@nao.cas.cn}
Zepeng Zheng (\begin{CJK*}{UTF8}{gbsn}郑泽鹏\ignorespacesafterend\end{CJK*}) \orcidlink{0009-0007-2280-1254}, $^{6}$
\newauthor
Long Wang (\begin{CJK*}{UTF8}{gbsn}王龙\ignorespacesafterend\end{CJK*}) \orcidlink{0000-0001-8713-0366}, $^{6,7}$
Xiaoying Pang (\begin{CJK*}{UTF8}{gbsn}庞晓莹\ignorespacesafterend\end{CJK*}) \orcidlink{0000-0003-3389-2263}, $^{8,9}$
Bo Zhang (\begin{CJK*}{UTF8}{gbsn}章博\ignorespacesafterend\end{CJK*}) \orcidlink{0000-0002-6434-7201}, $^{1}$
Hao Tian (\begin{CJK*}{UTF8}{gbsn}田浩\ignorespacesafterend\end{CJK*}) \orcidlink{0000-0003-3347-7596}, $^{1,4}$
\newauthor
Yang Xiang (\begin{CJK*}{UTF8}{gbsn}向阳\ignorespacesafterend\end{CJK*}) \orcidlink{0009-0007-9693-7963} $^{8}$
\\
$^{1}$National Astronomical Observatories, Chinese Academy of Sciences, Beijing 100101, People's Republic of China\\
$^{2}$School of Astronomy and Space Science, University of Chinese Academy of Sciences, Beijing 100049, People's Republic of China\\
$^{3}$INAF – Osservatorio di Astrofisica e Scienza dello Spazio di Bologna, via Piero Gobetti 93/3, 40129 Bologna, Italy\\
$^{4}$Institute for Frontiers in Astronomy and Astrophysics, Beijing Normal University, Beijing 100875, People’s Republic of China\\
$^{5}$Zhejiang Lab, Hangzhou, Zhejiang 311121, People’s Republic of China\\
$^{6}$School of Physics and Astronomy, Sun Yat-sen University, Daxue Road, Zhuhai, 519082, China\\
$^{7}$CSST Science Center for the Guangdong-Hong Kong-Macau Greater Bay Area, Zhuhai, 519082, China\\
$^{8}$Department of Physics, Xi'an Jiaotong-Liverpool University, Suzhou 215123, People’s Republic of China\\
$^{9}$Shanghai Key Laboratory for Astrophysics, Shanghai Normal University, Shanghai 200234, People’s Republic of China\\
}

\date{Accepted XXX. Received YYY; in original form ZZZ}

\pubyear{\the\year{}}

\begin{document}
\label{firstpage}
\pagerange{\pageref{firstpage}--\pageref{LastPage}}
\maketitle


\begin{abstract}
We present an IMF-aware N-body forward-modelling framework, in which the stellar initial mass function (IMF) is an explicit input, to quantify the membership selection functions of open clusters, namely completeness and contamination as functions of magnitude. Accurate derivation of the IMF requires correcting for these selection effects, which otherwise distort the inferred present-day mass function.
Our framework simulates clusters with IMF slopes $\alpha_2 = -2.7, -2.3, -1.9$, converts them into synthetic Gaia observations, adds mock field populations, and recovers members with {\hdbscan} to mimic observational membership selection. 
Applying this framework to 35 young open clusters (ages $\lesssim100$ Myr, distances $\lesssim600$ pc), we find declining completeness towards the faint limit; older, more expanded, and more diffuse models have lower completeness and higher contamination.
For a given cluster, models with flatter IMFs yield modestly but systematically lower global completeness and higher contamination, despite initial cluster masses calibrated to the same observed member count.
These models undergo more stellar evolution mass loss and expand more, resulting in lower projected stellar densities; their lower completeness is consistent with more diffuse clusters being harder to separate from field stars.
Corrections derived from a mismatched IMF largely recover the total star counts but leave a larger distortion in the shape of the present-day luminosity function than those derived with the correct IMF.
Because membership recovery is tied to a cluster's structural evolution and the specific membership algorithm used, our findings emphasize that selection functions should be inferred jointly with the IMF rather than assumed in advance.

\end{abstract}

\begin{keywords}
methods: data analysis -- catalogues -- astrometry -- stars: luminosity function, mass function -- open clusters and associations: general -- solar neighbourhood
\end{keywords}


\section{Introduction}
\label{sec:intro}

The stellar initial mass function (IMF), the distribution of stellar masses at birth, sets the luminosity, lifetime and feedback of a stellar population, and is one of the largest sources of systematic uncertainty in stellar and galaxy evolution models \citep{salpeter1955luminosity, hopkins2018dawes,kroupa2026initial}. It has been described by different mathematical forms, such as multi-part or tapered power laws \citep{kroupa2001variation,demarchi2010temporal} and log-normal distributions \citep{miller1979initial,chabrier2003galactic}. Beyond the choice of mathematical form, a long-standing question is whether the IMF is universal \citep{bastian2010universal} or varies systematically with star-forming density and metallicity \citep{marks2012evidence,li2023stellar}. Star clusters are well suited to testing these possibilities, because their members share a common age, distance and chemical composition, and clusters from different environments can be compared \citep{lada2003embedded, portegieszwart2010young, krumholz2019star}. Young nearby open clusters are particularly useful: when their ages are short compared with their relaxation times, mass segregation and the preferential loss of low-mass stars have had less time to alter their mass functions \citep{spitzer1987dynamical, baumgardt2003dynamical}, and \textit{Gaia} astrometry \citep{gaiacollaboration2023gaiab} allows their mass functions to be measured over a broad mass range \citep[e.g.][]{cordoni2023photometric}. Recent $N$-body modelling that accounts for dynamical evolution and unresolved binaries suggests tentative cluster-to-cluster scatter in the high-mass IMF slope \citep{hobart2026initial}. Whether this scatter reflects genuine IMF variations also depends on how well the incompleteness and field star contamination introduced by membership selection are quantified.

We therefore focus on this selection effect, which affects the mass function derived from any star counting approach. Incompleteness and contamination that are independent of magnitude would only rescale the mass function, whereas those that vary with magnitude, and hence with stellar mass, distort its shape. Faint stars have larger astrometric uncertainties \citep{lindegren2021gaia}, which broaden their observed distributions in proper motion and parallax. In addition, low mass stars are preferentially found in the outskirts of mass segregated clusters \citep{tarricq2022structural}, and many young clusters are observed to be expanding \citep{dellacroce2024young}, reducing the density contrast of their members against the field. Both effects make faint members harder to separate from field stars. Correcting the mass function of a cluster therefore requires the completeness $f_{\rm comp}(G)$, the fraction of true members that are recovered, and the contamination $f_{\rm cont}(G)$, the fraction of selected stars that are field stars, as functions of magnitude. We refer to these as the membership selection functions; they are specific to both the cluster and the procedure used to identify its members. 
They differ from the selection function of the \textit{Gaia} catalogue itself, which describes whether a star is included in the catalogue \citep{cantat-gaudin2023empirical}.

\textit{Gaia} has enabled homogeneous member lists for thousands of open clusters \citep{cantat-gaudin2018gaia, castro-ginard2020hunting, castro-ginard2022hunting, hunt2023improvinga, perren2023unified}, constructed with a range of membership algorithms \citep{campello2013densitybased, hunt2021improving, pera2021pyupmask, vangroeningen2023machinelearningbased}. The membership probabilities provided with these lists are specific to each algorithm and are not, in general, calibrated estimates of the completeness and contamination of the lists. \citet{hunt2024improving} estimated magnitude-dependent completeness for individual clusters and used it to correct their masses, but modelled member recovery from astrometric uncertainties alone, without field star contamination. \citet{hunt2026selection} went further: in addition to the probability that a cluster as a whole is detected, they illustrated both completeness and contamination at the membership level with injection--recovery tests, but their synthetic clusters share a fixed Kroupa IMF \citep{kroupa2001variation} and King spatial profiles \citep{king1962structure}, without dynamical evolution.

Dedicated mass function studies have typically mitigated these effects rather than directly correcting the selection functions of individual member stars. For example, some studies restrict their analysis to magnitude ranges where the survey is complete \citep{olivares2018seven}. Others select members conservatively and verify their completeness against deeper surveys \citep{hobart2026initial}, or use simulated clusters to demonstrate that their member lists are sufficiently complete and pure without correction \citep{jiang2024determination}. Alternatively, some studies avoid strict membership selection altogether by fitting field stars as a separate component in the colour--magnitude diagram \citep{li2025mimo, li2026evolution}. For existing member stars, completeness and contamination as functions of magnitude, including the effects of cluster evolution and structure, therefore remain largely unquantified.

Measuring these selection functions requires the true membership of every star, which is unknown for observed clusters but known by construction in simulations. Simulating a cluster, however, requires assumptions about its stellar population and structure. The IMF is particularly problematic: if the selection functions depend on it, corrections based on a fixed IMF would partly assume the very quantity they are meant to constrain.
The IMF can affect the selection functions directly, by setting how many members lie at each magnitude relative to field stars, and indirectly through the evolution of the cluster, since an IMF richer in massive stars leads to stronger stellar evolution mass loss and expansion \citep{chernoff1990evolution, kouwenhoven2014how, haghi2020lifetimes, wang2021impact}. Capturing this indirect effect requires evolving each simulated cluster from an initial state, rather than imposing a fixed present-day profile. In addition, the selection functions depend on the cluster, its surrounding field and the membership procedure \citep{hunt2021improving, hunt2024improving}, so those derived from generic simulated clusters cannot be transferred directly to a particular member list. 

$N$-body models of individual clusters have been used to infer their initial conditions and IMF \citep{baumgardt2023evidence, weis2025dynamical, hobart2026initial} and to calibrate their initial masses \citep{zheng2026binary}, but the simulated stars are compared with the observations without passing through the same membership selection as the observed members.
Deriving selection functions for a given member stars therefore requires cluster-specific $N$-body models with the IMF as an explicit input, whose simulated stars pass through the same membership selection as the observed members.

We construct such models with an IMF-aware $N$-body forward-modelling framework. For each cluster and each assumed IMF, we evolve an $N$-body model to the cluster age, observe it synthetically with the \textit{Gaia} uncertainties of that cluster and a mock field population, and identify its members with the same clustering algorithm as used for the observed list. The initial cluster mass is calibrated so that the number of recovered members matches the observed one, which fixes the initial mass for each assumed IMF. The completeness and contamination are then measured from the known origin of each simulated star, with their scatter estimated from independent realizations.
The framework can be applied to member lists from other algorithms by replacing only the membership identification step. We apply it to 35 nearby young open clusters with {\hdbscan} member lists \citep{hunt2023improvinga, hunt2024improving}, using three IMF slopes above $0.5\,M_\odot$ ($\alpha_2=-2.7$, $-2.3$, and $-1.9$), and examine how the selection functions vary with magnitude and between clusters, whether they depend on the assumed IMF, whether corrections derived with a mismatched IMF bias the recovered luminosity function, and how these differences relate to the evolution and present-day structure of the models.

The paper is organized as follows. Section~\ref{sec:observation_sample} describes the observational sample to which we apply the framework. Section~\ref{sec:method} introduces the IMF-aware $N$-body forward-modelling framework and details its construction. Section~\ref{sec:results} presents the magnitude-dependent and global completeness and contamination for different IMF slopes, their effect on the recovered present-day luminosity function, and their dependence on cluster properties. Section~\ref{sec:discussion} discusses the evolutionary and structural changes associated with the IMF dependence, the dependence on the membership algorithm, and the caveats of this work. Section~\ref{sec:conclusions} summarizes our conclusions.

\section{Observational sample}
\label{sec:observation_sample}
Our observational sample consists of 35 nearby young open clusters with distances $\lesssim 600$pc and ages $\lesssim 100$Myr.
The targets are initially selected by cross-matching the clusters studied by \citet{pang20223d} with the \textit{Gaia} DR3 catalogue of \citet{hunt2024improving}. Of the 37 matched objects, Alessi24 and BH164 are excluded because the mock-field data required for their simulations were unavailable.
All observational member lists and reference cluster properties adopted in this work are taken from \citet{hunt2023improvinga, hunt2024improving}. The identifications and adopted properties of the 35 clusters are listed in Table~\ref{tab:observation_properties} and described in Appendix~\ref{app:observation_properties}.

This work focuses on the demonstration of the method that can provide the IMF-aware model for the membership selection function, which plays an important role in mass function determination of a given star cluster, rather than on handling as many star clusters as possible. Therefore, the sample needs to contain sufficient clusters to demonstrate the method, and should also consist predominantly of young clusters, which are critical for future IMF measurement studies.

The observational member lists adopted in this work were obtained with {\hdbscan} in the all-sky search of \citet{hunt2023improvinga}. {\hdbscan}\footnote{\url{https://hdbscan.readthedocs.io/}} \citep[Hierarchical {\dbscan},][]{campello2013densitybased} combines the density-based approach of {\dbscan} \citep[Density-Based Spatial Clustering of Applications with Noise,][]{ester1996densitybased} with hierarchical clustering to effectively handle datasets of varying densities. To search for clusters closer than 750~pc, \citet{hunt2023improvinga} divided the sky into 192 level-2 HEALPix (Hierarchical Equal Area isoLatitude Pixelation) pixels and selected sources with $\varpi>1$~mas. Each pixel and its eight nearest neighbours formed a field of approximately $44^\circ\times44^\circ$, with adjacent fields overlapping. For clusters closer than 150~pc, they instead used a single all-sky field containing stars with photo-geometric distances below 250~pc (see their Section~2.3). Within each HEALPix field, positions, proper motions, and parallaxes were transformed to a coordinate frame centred on the field and rescaled so that all five dimensions had the same median and interquartile range. For the all-sky field, positions and parallaxes were converted to Cartesian coordinates using photo-geometric distances, and proper motions to tangential velocities (their Section~3.1). They adopted the same {\hdbscan} hyperparameter values as in their previous study \citep{hunt2021improving}, with $m_{\rm clSize}\in\{10,20,40,80\}$ and $m_{\rm Pts}=10$. Detections were required to exceed $3\sigma$ in the cluster significance test before the results from different hyperparameter runs and fields were combined and duplicate detections removed. During this merging, only clusters closer than 175~pc were retained from the all-sky search, whereas those from the level-2 HEALPix search were retained only between 150 and 750~pc. The overlap allowed the better detection to be selected for clusters near the boundary (see their Section~3.2). The minimum cluster size $m_{\rm clSize}$ adopted for each object is recorded in the online cluster catalogue, as described in their Table~A.1. We use the same {\hdbscan} algorithm to recover members in our forward models, as described in Section~\ref{sec:membership_recovery}.

\section{Method}
\label{sec:method}
\subsection{Forward-Modelling Framework}
\label{sec:forward_framework}
\begin{figure*}
\includegraphics[width=\textwidth]{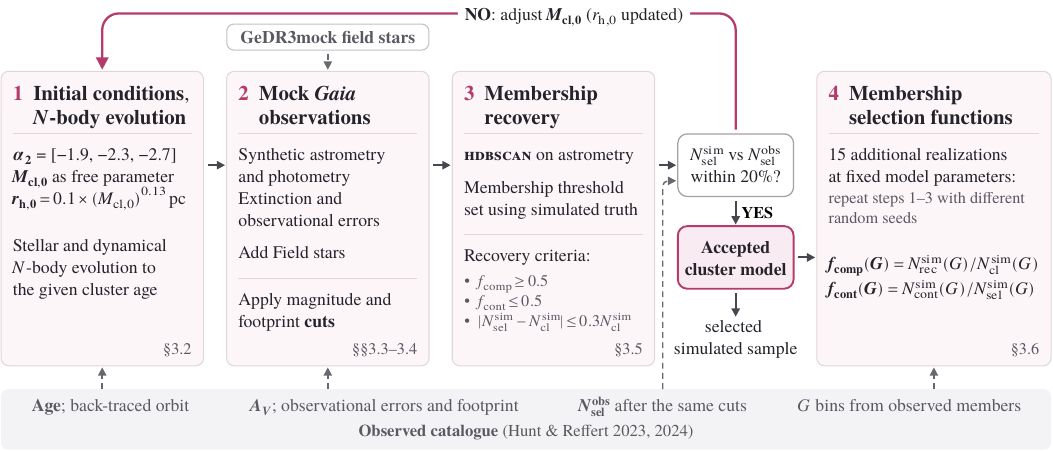}
\centering
\caption{
    Forward-modelling framework for a given cluster and IMF slope $\alpha_2$ above $0.5\ M_\odot$. Steps 1--3 produce a simulated sample from an initial cluster mass $M_{\rm cl,0}$, which is adjusted until the number of selected simulated sample $N_{\mathrm{sel}}^{\mathrm{sim}}$, agrees with the observed member count $N_{\mathrm{sel}}^{\mathrm{obs}}$, to within 20 per cent. Step 4 measures the membership selection functions for the  accepted model. The bottom row lists the inputs taken from the observed catalogue \citep{hunt2023improvinga,hunt2024improving}; section numbers refer to the text.
}
\label{fig:framework}
\end{figure*}

We construct an IMF-aware forward model that follows each observed cluster from an assumed initial state to a present-day membership catalogue, as shown in Fig.~\ref{fig:framework}. The model combines $N$-body simulation (step~1), synthetic \textit{Gaia} observations with a mock field population (step~2), and {\hdbscan} membership recovery (step~3). The IMF is an explicit input: for each IMF slope $\alpha_2$, we generate a new set of stellar masses and follow the stellar and dynamical evolution of the resulting cluster in a separate $N$-body model. The output is a simulated counterpart of the observed member list, subject to the same magnitude and footprint cuts as the observed sample. As in \citet{zheng2026binary}, we calibrate $M_{\rm cl,0}$ by matching simulated and observed star counts, but after membership recovery rather than within the tidal radius, so that both counts are subject to incompleteness and contamination. We repeat steps~1--3 with trial values of $M_{\rm cl,0}$, updating $r_{\rm h,0}$ through the adopted mass--radius relation (Section~\ref{sec:imf_nbody}), until the selected simulated and observed sample counts agree to within 20 per cent (Section~\ref{sec:membership_calibration}). In step~4, we use the known origin of each simulated star in the accepted model to measure the membership selection functions, namely completeness and contamination as functions of $G$ magnitude, and repeat the calculation across independent realizations to estimate their scatter.

For a given $\alpha_2$, the same forward model, which follows the cluster from its initial state to the present day, thus serves both to calibrate the initial mass and to derive the selection functions. For each cluster, we obtain an accepted model for each $\alpha_2$, all matching the same observed member count within the adopted tolerance. Comparing their selection functions allows us to examine whether and how they vary with the assumed IMF for that cluster.

\subsection{Initial Conditions and N-body Evolution}
\label{sec:imf_nbody}

We generate the initial cluster models with a modified version of \textsc{mcluster}\footnote{\url{https://github.com/yueyue-shen/mcluster}} \citep{wang2019complete, kupper2011mass}. We adopt a continuous, two-part power-law IMF \citep{kroupa2001variation},
\begin{equation}
 \xi(m)=\frac{{\rm d}N}{{\rm d}m}=k
 \begin{cases}
  (m/0.5\,M_\odot)^{-1.3}, & 0.08\leq m/M_\odot<0.5,\\
  (m/0.5\,M_\odot)^{\alpha_2}, & 0.5\leq m/M_\odot\leq150,
 \end{cases}
 \label{eq:imf}
\end{equation}
where $k$ is the normalization constant. We consider $\alpha_2=-1.9$, $-2.3$, and $-2.7$ to compare membership selection functions under different IMFs. We assign stellar masses by optimal sampling \citep{kroupa2013stellar} to reduce stochastic variations in the evolution of low-mass clusters \citep{wang2021impact}. For each $M_{\rm cl,0}$ and $\alpha_2$, we recalculate the discrete mass spectrum and maximum initial stellar mass $m_{\rm max}$. In our implementation, $k$ and $m_{\rm max}$ jointly satisfy 
\begin{equation}
 \int_{m_{\rm max}}^{150\,M_\odot}\xi(m)\,{\rm d}m=1,
 \qquad
 \int_{0.08\,M_\odot}^{m_{\rm max}}m\xi(m)\,{\rm d}m=M_{\rm cl,0}.
 \label{eq:maximum_stellar_mass}
\end{equation}
Here $150\,M_\odot$ is the IMF upper limit, while $m_{\rm max}$ is the mass of the most massive star assigned to the model.

We calibrate the initial cluster mass $M_{\rm cl,0}$ separately for each $\alpha_2$ (Section~\ref{sec:membership_calibration}) and set the initial half-mass radius using the mass--radius relation of \citet{marks2012inverse},
$r_{\rm h,0}=0.1\times(M_{\rm cl,0}/M_\odot)^{0.13}$~pc.
For our calibrated masses of $M_{\rm cl,0}\approx25$--$1800\,M_\odot$, this gives $r_{\rm h,0}\approx0.15$--$0.26$~pc. Models adopting this relation may remain more compact at the present day than the observed clusters \citep{zheng2026binary}. We nevertheless retain this relation to limit computational cost and avoid parameter degeneracies that would be difficult to resolve if $r_{\rm h,0}$ were varied independently alongside $M_{\rm cl,0}$ and $\alpha_2$.  The clusters are initialized as King models with $W_0=5$ \citep{king1966structure}, scaled to $r_{\rm h,0}$ and in virial equilibrium, without primordial mass segregation, fractal substructure, or primordial binaries.

We evolve the clusters with the $N$-body code {\petar}\footnote{\url{https://github.com/lwang-astro/PeTar}} \citep{wang2020petar}, which combines the particle-particle particle tree (P$^{3}$T) method \citep{oshino2011particle} with slow-down algorithmic regularization (SDAR; \citealp{wang2020slowdown}) and is parallelized with the Framework for Developing Particle Simulators (FDPS; \citealp{iwasawa2016implementation,iwasawa2020accelerated}). Stellar and binary evolution, including wind mass loss, stellar-type transitions, mass transfer, and mergers, is followed with the updated \textsc{sse}/\textsc{bse} codes \citep{hurley2000comprehensive,hurley2002evolution,banerjee2020bse}, and the Galactic potential is included through {\galpy} \citep{bovy2015galpy}.

We adopt $Z=0.02$ for stellar evolution and \texttt{MWPotential2014} for the Galactic potential. For each cluster, a centre-of-mass particle is integrated backwards from the present-day Galactocentric position and velocity derived from the catalogue of \citet{hunt2024improving}. The backward integration time is the catalogue median age rounded down to the nearest Myr (Appendix~\ref{app:observation_properties}). The $N$-body models for all mass and IMF trials of the cluster start from the resulting position and velocity and are evolved forward over this interval to the present day. In the final snapshot, we retain all particles, including remnants and escapers still in the simulation, without any radial or bound-membership cut. This snapshot is the input to the synthetic \textit{Gaia} observations described in Section~\ref{sec:synthetic_gaia}.

\subsection{Synthetic Gaia Observations and Footprint}
\label{sec:synthetic_gaia}

We use \textsc{galevnb} \citep{pang2016galevnb} to compute absolute $G$, $G_{\rm BP}$, and $G_{\rm RP}$ magnitudes only for particles with \texttt{star.type}~$< 10$ (i.e. main-sequence, Hertzsprung-gap, first-giant-branch, core-helium-burning, and asymptotic-giant-branch stars, and naked helium stars). We assign all stars in each cluster the catalogue median $A_V$ (\texttt{AV50}) from \citet{hunt2024improving} and convert it into extinction in each band using the corrected polynomial coefficients of \citet{monteiro2024correction}, with extrapolation where necessary. We then obtain noiseless apparent magnitudes by adding the distance modulus for each star's simulated heliocentric distance and the corresponding band extinction to its absolute magnitudes.

To reproduce the measurement uncertainties in each cluster, we derive empirical magnitude--uncertainty relations from all of its catalogue members \citep{hunt2024improving}, rather than adopting general \textit{Gaia} DR3 astrometric uncertainty models \citep{lindegren2021gaia}. Using locally weighted scatterplot smoothing (LOWESS), we fit the logarithmic uncertainties of the parallax and proper motions as functions of $G$, and the logarithmic uncertainty in each photometric band as a function of the magnitude in that band. For each mock star, we interpolate these relations at its noiseless magnitudes and add independent zero-mean Gaussian noise with the resulting standard deviations to its parallax, proper motions, and three magnitudes. We then retain sources with $4\leq G\leq19$~mag.

We further restrict the sources to a footprint in sky position, parallax, and proper motion, defined from observed members of each cluster and fixed throughout the trials. The footprint radius is defined as $R_{\rm f}=\max(3r_{95},\,r_{95}+20~{\rm pc})$, where $r_{95}$ encloses 95 per cent of the observed members in projection at the catalogue distance $d_0$ (\texttt{dist50}). The corresponding angular radius $\theta_{\rm f}=\arctan(R_{\rm f}/d_0)$ is limited to $0.5^\circ$--$10^\circ$. The proper motion cut is an ellipse centred on the catalogue proper motion, with semi-axes $2\Delta\mu$ in each component, where $\Delta\mu=\max(1.5\mu_{95},\,\mu_{10},\,3e_\mu)$, limited to 0.3--5~mas\,yr$^{-1}$, and $\mu_{10}$ corresponds to a transverse speed of 10~km\,s$^{-1}$ at $d_0$. The parallax window is centred on the cluster parallax $\varpi_0$, with half-width $\Delta\varpi=\max(1.5\varpi_{95},\,3e_\varpi,\,\varpi_0R_{\rm f}/d_0,\,0.05\varpi_0)$, limited to 0.05--1~mas. 
Here $\mu_{95}$ and $\varpi_{95}$ are the 95th percentiles of the members' absolute offsets from the respective window centres, and $e_\mu$ and $e_\varpi$ are their median reported uncertainties, the term $\varpi_0R_{\rm f}/d_0$ allows for the cluster's line-of-sight depth. We retain stars within the angular radius, parallax window, and proper-motion ellipse:
\begin{equation}
\begin{aligned}
&\theta\leq\theta_{\rm f},\qquad
|\varpi-\varpi_0|\leq\Delta\varpi,\\
&\left(\frac{\mu_{\alpha *}-\mu_{\alpha *,0}}
{\Delta\mu_{\alpha *}}\right)^2
+\left(\frac{\mu_\delta-\mu_{\delta,0}}
{\Delta\mu_\delta}\right)^2<4.
\end{aligned}
\label{eq:footprint_selection}
\end{equation}
Here, $\theta$ is the angular separation from the catalogue centre, and $\mu_{\alpha *,0}$ and $\mu_{\delta,0}$ are the catalogue proper motions.

We apply the same $4\leq G\leq19$~mag and footprint cuts to the observed members and to the simulated cluster stars combined with mock field stars (Section~\ref{sec:mock_field}), so that both counts refer to the same selection region. The selected simulated sample is then passed to {\hdbscan} for membership identification (Section~\ref{sec:membership_calibration}).

\subsection{Mock Field Stars}
\label{sec:mock_field}

With the synthetic cluster observations prepared, we add stars from the Gaia Early DR3 Mock Stellar Catalogue (GeDR3mock)\footnote{\url{https://dc.g-vo.org/tableinfo/gedr3mock.main}} \citep{rybizki2020gaia}. This catalogue was generated with {\galaxia} \citep{sharma2011galaxia}, combining a Besan\c{c}on-like Galactic model with 3D extinction maps and stellar evolutionary models.
It models the selection effects of the \textit{Gaia} survey and provides \textit{Gaia}-like uncertainty estimates.

For each cluster, we download field stars from a region with the same centre as the footprint and twice its angular radius $\theta_{\rm f}$, proper-motion scales $\Delta\mu$, and parallax half-width $\Delta\varpi$. Besides, we also apply additional quality criteria as described in Appendix~\ref{app:mock_field_query}, those criteria are used only for download the field population, whereas the $4\leq G\leq19$~mag and footprint cuts (Section~\ref{sec:synthetic_gaia}) are applied afterwards to both populations. We retain the same field sample for each cluster throughout the calibration trials, providing a common background for the recovery comparison.

\subsection{Membership Recovery and Initial-Mass Calibration}
\label{sec:membership_calibration}
\subsubsection{Membership recovery with {\hdbscan}}
\label{sec:membership_recovery}

We adopt {\hdbscan} \citep{campello2013densitybased} because it was used to construct the homogeneous observational catalogue of \citet{hunt2023improvinga,hunt2024improving} and is efficient for repeated forward calculations. The inputs are $(\mu_{\alpha *},\mu_\delta,l,b,\varpi)$, with each dimension standardized by its mean and standard deviation.
The main hyperparameters are the minimum cluster size $m_{\rm clSize}$ and the number of samples in a neighbourhood $m_{\rm Pts}$. We fix $m_{\rm clSize}=10$ and $m_{\rm Pts}=5$ and use excess of mass (\texttt{eom}) cluster selection, keeping recovery consistent across trials and limiting repeated tuning; the rationale is given in Appendix~\ref{app:hdbscan_fix}.
These settings differ from those used in the all-sky search of \citet{hunt2023improvinga} because our footprint selection and subsequent data standardization also differ. Following their procedure more strictly would allow a more rigorous comparison between the simulated and observed member samples. Our aim, however, is to develop an IMF-aware forward-modelling framework and derive IMF-aware selection functions, rather than reproduce the all-sky search.

\begin{figure*}
\includegraphics[width=0.9\textwidth]{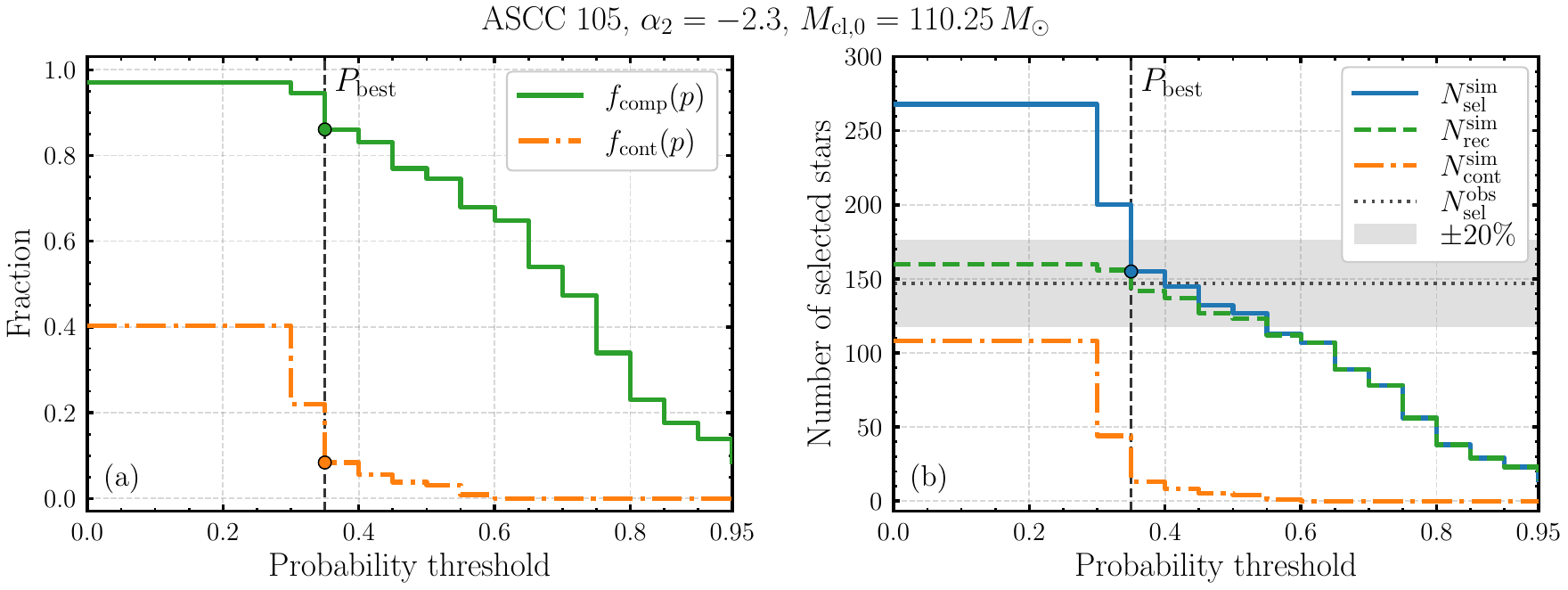}
\centering
\caption{
    Probability threshold selection and star count comparison for ASCC~105, with $\alpha_2=-2.3$ and $M_{\rm cl,0}=110.25\,M_\odot$.
    (a) Completeness and contamination as functions of the {\hdbscan} probability threshold $p$. The vertical dashed lines mark $P_{\rm best}=0.35$, which maximizes $f_{\rm comp}(p)-f_{\rm cont}(p)$; circles mark the corresponding values.
    (b) Counts of all selected sources $N_{\mathrm{sel}}^{\mathrm{sim}}$, correctly recovered members $N_{\mathrm{rec}}^{\mathrm{sim}}$, and field contaminants $N_{\mathrm{cont}}^{\mathrm{sim}}$. The dotted horizontal line marks the observed count $N_{\mathrm{sel}}^{\mathrm{obs}}=147$, and the shaded band shows its $\pm20$ per cent acceptance interval.
    At $P_{\rm best}$, $N_{\mathrm{sel}}^{\mathrm{sim}}=155$ satisfies the star count criterion for mass calibration.
}
\label{fig:Pbest_converage_example}
\end{figure*}

From the {\hdbscan} results obtained with the fixed settings, we identify the simulated target cluster as the non-noise label whose centroid is nearest to the observed cluster centre. Each source in this label receives a membership strength between 0 and 1, reported as a "probability". It is important to note that these values measure relative assignment strength rather than statistical membership probability and cannot be used to weight star counts; retaining all sources in the label would introduce disproportionately high field contamination. A suitable probability threshold must therefore be chosen: only sources that pass it are treated as simulated members in the final count comparison.
The best threshold $P_{\rm best}$ is determined by the maximum difference between membership completeness and contamination:
\begin{equation}
 P_{\rm best}=\operatorname*{arg\,max}_{p\in\{0,\,0.05,\,\ldots,\,0.95\}}\left[f_{\rm comp}(p)-f_{\rm cont}(p)\right],
 \label{eq:best_threshold}
\end{equation}
where selecting sources in the target label with membership strength of at least $p$ gives the completeness $f_{\rm comp}(p)$ and contamination $f_{\rm cont}(p)$ (equation~\ref{eq:magnitude_selection_functions}). $P_{\rm best}$ is determined from simulated truth. Figure~\ref{fig:Pbest_converage_example}(a) illustrates this threshold selection for ASCC~105. This threshold is crucial because it controls the trade-off between completeness and contamination. It also changes the selected member count, which affects the convergence of mass calibration.

At $P_{\rm best}$, we consider a simulated cluster successfully recovered if
\begin{equation}
 \begin{cases}
  f_{\rm comp}(P_{\rm best})\geq0.5,\\
  f_{\rm cont}(P_{\rm best})\leq0.5,\\
  \left|N_{\mathrm{sel}}^{\mathrm{sim}}-N_{\mathrm{cl}}^{\mathrm{sim}}\right|\leq0.3\,N_{\mathrm{cl}}^{\mathrm{sim}},
 \end{cases}
 \label{eq:recovery_criteria}
\end{equation}
where $N_{\mathrm{sel}}^{\mathrm{sim}}$ counts all selected sources, both correctly recovered members and field contaminants, while $N_{\mathrm{cl}}^{\mathrm{sim}}$ counts all cluster-origin particles passing the photometric and footprint cuts. We test convergence of the calibration only for recovered trials.

\subsubsection{Initial cluster mass calibration}
\label{sec:mass_calibration}
If the simulated cluster is recovered, we calibrate the trial against $N_{\mathrm{sel}}^{\mathrm{obs}}$, the observed member count after the same magnitude and footprint cuts, accepting differences of at most 20 per cent of this count. As illustrated in Fig.~\ref{fig:Pbest_converage_example}(b) for ASCC~105, the selected count at $P_{\rm best}$ falls within this acceptance interval. To find a mass that meets this criterion, we start from $M_{\rm cl,0}=0.75 \times N_{\mathrm{sel}}^{\mathrm{obs}}\,M_\odot$, following \citet{zheng2026binary} and \citet{marks2012inverse}, and at each trial mass repeat the forward calculation, recovery, and threshold selection. We first search for a mass interval that brackets the observed count by doubling the trial mass when too few stars are selected, or halving it when too many are selected. We then bisect the bracketed mass interval; this assumes that $N_{\mathrm{sel}}^{\mathrm{sim}}$ generally increases with initial mass. Convergence is typically reached after a few simulations, although changes in the {\hdbscan} recovery between trials can break this trend and prevent convergence.

If recovery fails, we instead compare counts after only the magnitude cut $4\leq G\leq19$: $N_{\mathrm{mag}}^{\mathrm{sim}}$ for simulated cluster stars with synthetic photometry, and $N_{\mathrm{mag}}^{\mathrm{obs}}$ for observed members. If $N_{\mathrm{mag}}^{\mathrm{sim}}<N_{\mathrm{mag}}^{\mathrm{obs}}$, the trial mass is too low and sets a lower bound for the bracketing interval, so we continue at higher masses; otherwise, we stop because the search direction is ambiguous. Such trials guide the search but are never accepted as calibrated models.

If mass calibration fails to converge with the fixed {\hdbscan} settings, we repeat it with a search over 28 hyperparameter combinations at each trial mass: $m_{\rm clSize}\in\{10,15,20,30,40,50,80\}$, $m_{\rm Pts}\in\{5,10\}$, and \texttt{eom} or \texttt{leaf} selection. In practice, only a small fraction of our sample required this additional search. For each combination, the target is the label with the most true cluster stars. We retain the combination maximizing the area under the completeness--contamination curve, then reselect $P_{\rm best}$ using equation~(\ref{eq:best_threshold}). The search thus uses simulated truth for both target and hyperparameter selection; details are given in Appendix~\ref{app:hdbscan_search}.

\subsection{Membership Selection Functions}
\label{sec:selection_functions}

The selection functions compare the simulated samples before and after {\hdbscan} membership recovery (Section~\ref{sec:membership_recovery}). Before recovery, $N_{\mathrm{all}}^{\mathrm{sim}}$ counts all cluster-origin particles in the final snapshot (Section~\ref{sec:imf_nbody}), including remnants and escapers; $N_{\mathrm{mag}}^{\mathrm{sim}}$ counts those eligible for synthetic photometry and within the adopted magnitude limits; and $N_{\mathrm{cl}}^{\mathrm{sim}}$ counts the subset also inside the footprint, i.e. the cluster stars in the {\hdbscan} input. At $P_{\rm best}$, the selected sample comprises correctly recovered cluster members and field contaminants, with counts $N_{\mathrm{rec}}^{\mathrm{sim}}$ and $N_{\mathrm{cont}}^{\mathrm{sim}}$, respectively, giving $N_{\mathrm{sel}}^{\mathrm{sim}}=N_{\mathrm{rec}}^{\mathrm{sim}}+N_{\mathrm{cont}}^{\mathrm{sim}}$.

We define the membership selection functions as the magnitude-dependent completeness $f_{\rm comp}(G)$ and contamination $f_{\rm cont}(G)$:
\begin{equation}
 f_{\rm comp}(G)=\frac{N_{\mathrm{rec}}^{\mathrm{sim}}(G)}{N_{\mathrm{cl}}^{\mathrm{sim}}(G)},\qquad
 f_{\rm cont}(G)=\frac{N_{\mathrm{cont}}^{\mathrm{sim}}(G)}{N_{\mathrm{sel}}^{\mathrm{sim}}(G)}.
 \label{eq:magnitude_selection_functions}
\end{equation}
We use equal-frequency magnitude bins defined from the observed sample to accommodate their non-uniform magnitude distribution. Following the Rice rule, we set the number of bins to $k=\lceil2(N_{\mathrm{sel}}^{\mathrm{obs}})^{1/3}\rceil$. The global completeness and contamination are computed using the same ratios, with counts summed over the full adopted magnitude interval.

To improve the statistical robustness of our estimates, we perform 15 additional simulations for each accepted model, using different random seeds to generate the initial stellar masses, positions, and velocities while keeping the calibrated total mass and associated radius, IMF, age, observational-error model and noise seed, field population, and footprint fixed. Each realization is evolved independently and subjected to membership recovery, with $P_{\rm best}$ reselected but no mass recalibration. For models calibrated through a hyperparameter search, these repeats use the reference model's final hyperparameters and target identification rule without further searches.

From the accepted model and 15 additional realizations, we retain only those meeting the global recovery criteria (equation~(\ref{eq:recovery_criteria})) and report the median and 16th--84th percentile range in each magnitude bin, along with the number of valid realizations. These summaries describe scatter among successfully recovered realizations under fixed model and field assumptions, rather than uncertainty in the median or systematic uncertainties.

\section{Results}
\label{sec:results}
\subsection{Magnitude-dependent completeness and contamination}
\label{sec:mag-depend}
\begin{figure*}
\includegraphics[width=\textwidth]{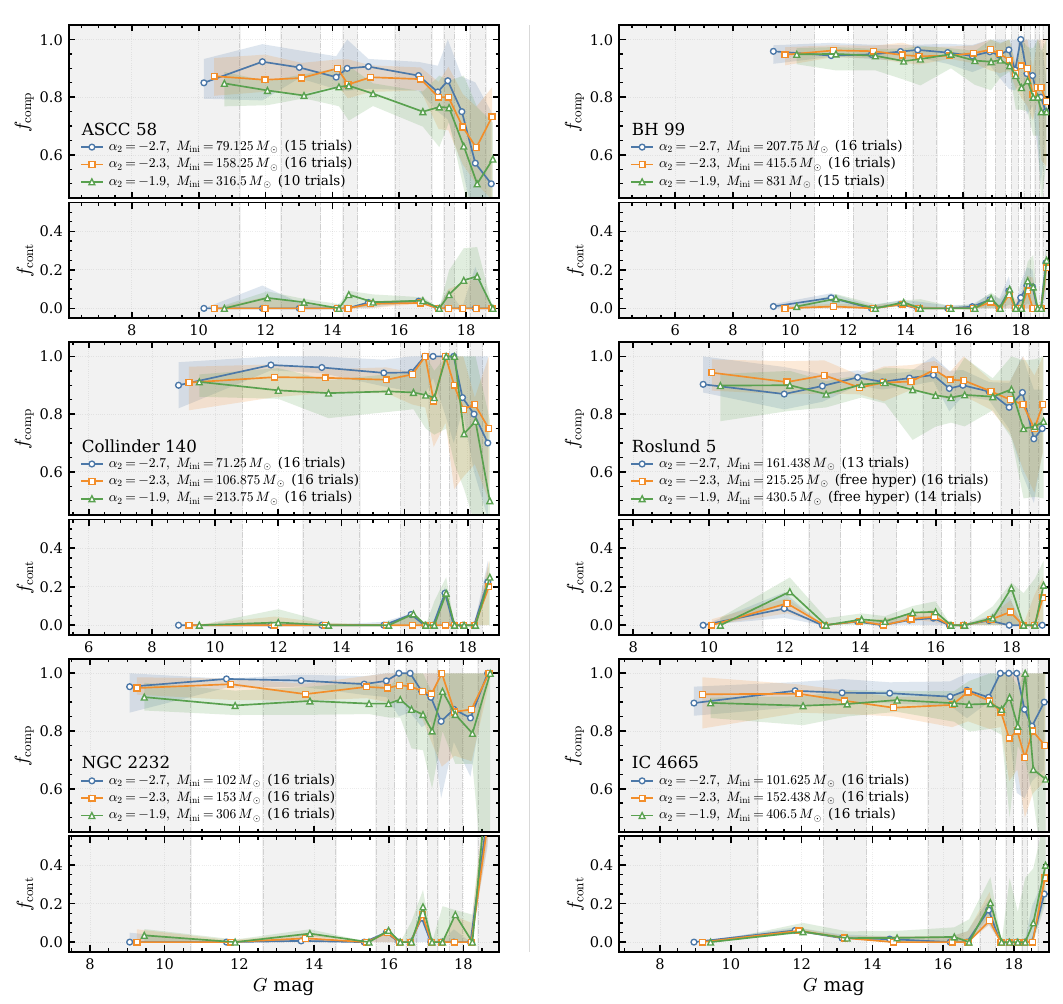}
\centering
\caption{
    Selection functions for six example open clusters. Upper and lower panels show completeness $f_{\rm comp}$ and contamination $f_{\rm cont}$ versus $G$ magnitude, respectively. Blue circles, orange squares, and green triangles denote $\alpha_2=-2.7$, $-2.3$, and $-1.9$, respectively. Curves and shading show the medians and 16th--84th percentile ranges across realizations. Alternating grey and white vertical bands mark the $G$-magnitude bins. Legends give the initial cluster masses and numbers of contributing realizations; "free hyper" denotes models with optimized {\hdbscan} hyperparameters.
}
\label{fig:selection_IMF}
\end{figure*}

We attempted to calibrate the initial mass of each of the 35 clusters for 3 assumed IMF slopes, $\alpha_2=-2.7$, $-2.3$, and $-1.9$, giving 105 combinations. The mass calibration converged for 97 combinations: 92 with the fixed {\hdbscan} hyperparameters and 5 through the supplementary hyperparameter search. The remaining 8 combinations failed the membership recovery criteria even after this search. The accepted models used in the following selection function analysis therefore cover 34 clusters, all except IC~2391, for which none of the three combinations satisfied the recovery criteria. Results are available for all three IMF slopes for 30 clusters, two slopes for three clusters, and one slope for IC~2602.

Across the full sample shown in Fig.~\ref{fig:selection_IMF} and Fig.~\ref{fig:selection_full}, median completeness typically lies between $0.8$ and $1$ at $G\lesssim17$, while median contamination is generally below $0.1$. Towards the faint limit, completeness declines to $\sim0.5$--$0.8$ in many clusters, although others retain higher values. Contamination varies substantially between clusters, remaining below $0.1$ in many faint bins but exceeding $0.5$ in the faintest bins of several clusters, such as NGC~2232. These magnitude-dependent selection effects make corrections for incompleteness and contamination essential when deriving mass functions. 

Since the simulated clusters used to estimate these corrections depend on the assumed IMF, we examine whether and how the selection functions differ among the three values of $\alpha_2$. For about half of the clusters, the median completeness decreases systematically from $\alpha_2=-2.7$ to $-1.9$ over most of the magnitude range, as illustrated by ASCC~58, Collinder~140, and NGC~2232 in the left-hand column of Fig.~\ref{fig:selection_IMF}. However, the separations between the medians, typically $\lesssim0.1$, are comparable to or smaller than the 16th--84th percentile ranges across realizations, so the realization scatter does not allow us to establish whether this ordering is statistically significant. Moreover, even in these clusters, the shapes of the selection functions do not always follow the same trend for different $\alpha_2$: at $G\gtrsim17$, the curves for the three slopes cross, and the decline in completeness towards the faint limit differs between slopes. For example, in the faintest bin of Collinder~140, the completeness falls to 0.5 for $\alpha_2=-1.9$ but remains at 0.70--0.75 for the other two slopes. For the remaining clusters, the selection functions show no systematic dependence on $\alpha_2$, or the scatter is too large to identify one: the medians for the three slopes either nearly coincide, as for BH~99, or cross without a consistent order, as for Roslund~5 and IC~4665 in the right-hand column of Fig.~\ref{fig:selection_IMF}.

\begin{figure*}
\includegraphics[width=0.9\textwidth]{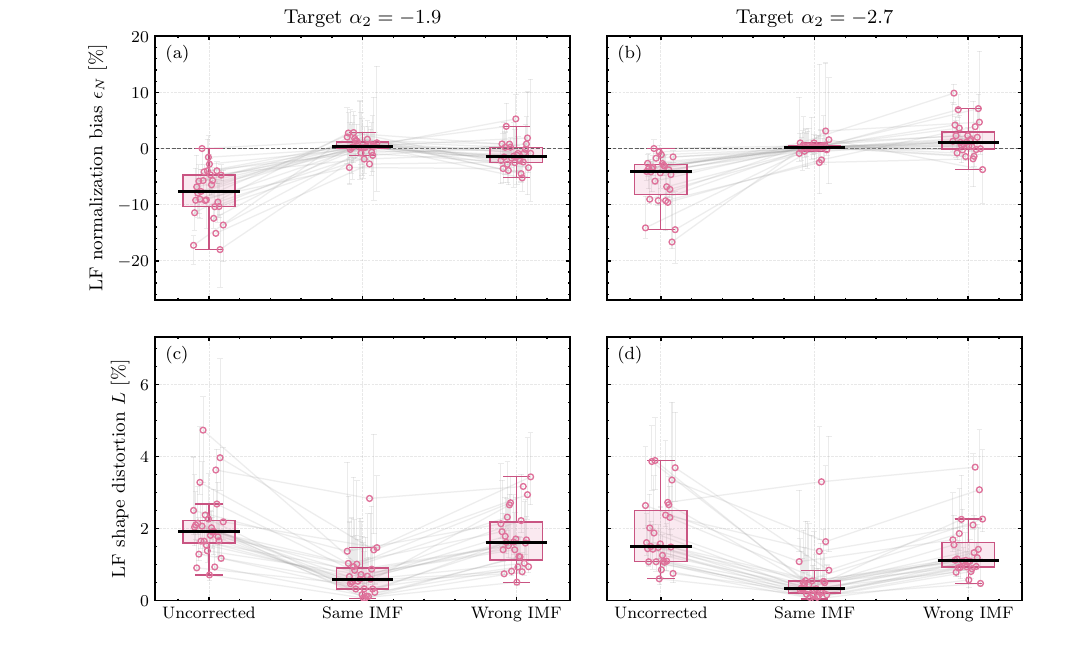}
\centering
\caption{
    Normalization bias, $\epsilon_N$ (top), and shape distortion, $L$ (bottom), of the luminosity function (LF) derived from the selected sample before and after selection-function correction, measured relative to the true present-day LF (PDLF) over the full original magnitude-bin range and expressed in per cent. The left and right columns correspond to target models with $\alpha_2=-1.9$ and $-2.7$, respectively. Uncorrected denotes no correction; Same IMF and Wrong IMF denote corrections derived from independent realizations with the target $\alpha_2$ and with $\alpha_2=-2.3$, respectively. Pink open circles show individual clusters, and grey lines connect the same cluster across the three cases. Grey error bars show 95 per cent bootstrap intervals. Boxes show the interquartile ranges across clusters, with black lines marking the medians and whiskers extending to 1.5 times the interquartile range.
}
\label{fig:PDLF_count_shape}
\end{figure*}

To assess whether these differences matter once the corrections are applied, we test how well they recover the true present-day luminosity function (PDLF) when derived with the correct IMF and with a different one. Because the selection functions are defined in magnitude bins, we perform this test on the luminosity function (LF) itself, without a mass--luminosity conversion. We treat each realization of the models with $\alpha_2=-1.9$ and $-2.7$ in turn as a target and correct its selected sample using the selection functions of other realizations of the same cluster. We compare three cases: no correction; same-IMF corrections, which use the other realizations of the target model; and wrong-IMF corrections, which use the realizations of the $\alpha_2=-2.3$ model. We use only realizations with valid counts in all magnitude bins, so that every LF is compared with the PDLF over the same full magnitude range.

In each magnitude bin $b$, let $N_b^{\rm true}$ be the true number of cluster stars in the target and $N_b$ the recovered number. Without correction, $N_b$ is the selected count, $N_b^{\rm sel}$; with correction, $N_b=N_b^{\rm sel}(1-f_{\rm cont})/f_{\rm comp}$, with $f_{\rm comp}$ and $f_{\rm cont}$ taken from one correction realization. We quantify the recovery with two diagnostics, evaluated for each pair of target and correction realizations and summarized for each cluster by the median over correction realizations and then over targets. 
The normalization bias,
\begin{equation}
 \epsilon_N=\frac{\sum_b N_b}{\sum_b N_b^{\rm true}}-1,
 \label{eq:normalization_bias}
\end{equation}
is the signed relative error in the total number of stars, with positive values indicating overestimated counts.
The shape distortion $L$ compares the LFs after normalizing each to unit total, so it is insensitive to the total number of stars,
\begin{equation}
 L=\frac{1}{2}\sum_b\left|\operatorname{median}\left(\frac{N_b}{\sum_{b'}N_{b'}}-\frac{N_b^{\rm true}}{\sum_{b'}N_{b'}^{\rm true}}\right)\right|,
 \label{eq:shape_distortion}
\end{equation}
where the median is taken in each bin before the absolute value. $L$ therefore measures the systematic change in LF shape, rather than the random scatter between realizations. For each cluster, we obtain 95 per cent intervals on $\epsilon_N$ and $L$ from 5000 bootstrap resamples of its target realizations, with the corrections kept fixed.

Fig.~\ref{fig:PDLF_count_shape} shows that the uncorrected LFs are biased in both normalization and shape. Without correction, $\epsilon_N$ is negative for almost all clusters, with medians of $-8$ and $-4$ per cent for targets with $\alpha_2=-1.9$ and $-2.7$, respectively, and the median $L$ is $2$ and $1.5$ per cent. Same-IMF corrections reduce both biases: the median $\epsilon_N$ becomes close to zero, the median $L$ decreases to $0.5$ and $0.3$ per cent, and the scatter between clusters decreases. Wrong-IMF corrections, which use the $\alpha_2=-2.3$ models, leave median $\epsilon_N$ values of $-1.4$ per cent for targets with $\alpha_2=-1.9$ and $+1.2$ per cent for those with $\alpha_2=-2.7$. Most of the normalization bias is therefore still removed, but the total counts remain underestimated when the target IMF is flatter than $\alpha_2=-2.3$ and overestimated when it is steeper. The shape distortion, however, is reduced much less. Fig.~\ref{fig:PDLF_count_shape}(c) and (d) show that wrong-IMF corrections leave median $L$ values of $1.6$ and $1.1$ per cent, about three times the same-IMF values and only modestly below the uncorrected ones. Besides this higher median, $L$ also varies more between clusters than with same-IMF corrections, suggesting that clusters respond differently to the IMF mismatch. Overall, corrections derived with the correct IMF largely recover both the normalization and the shape of the LF, whereas those derived with a mismatched IMF bring the total number of stars close to its true value but still leave a shape distortion.

\subsection{Global completeness and contamination}
\label{sec:global}
 \begin{figure*}
\includegraphics[width=0.9\textwidth]{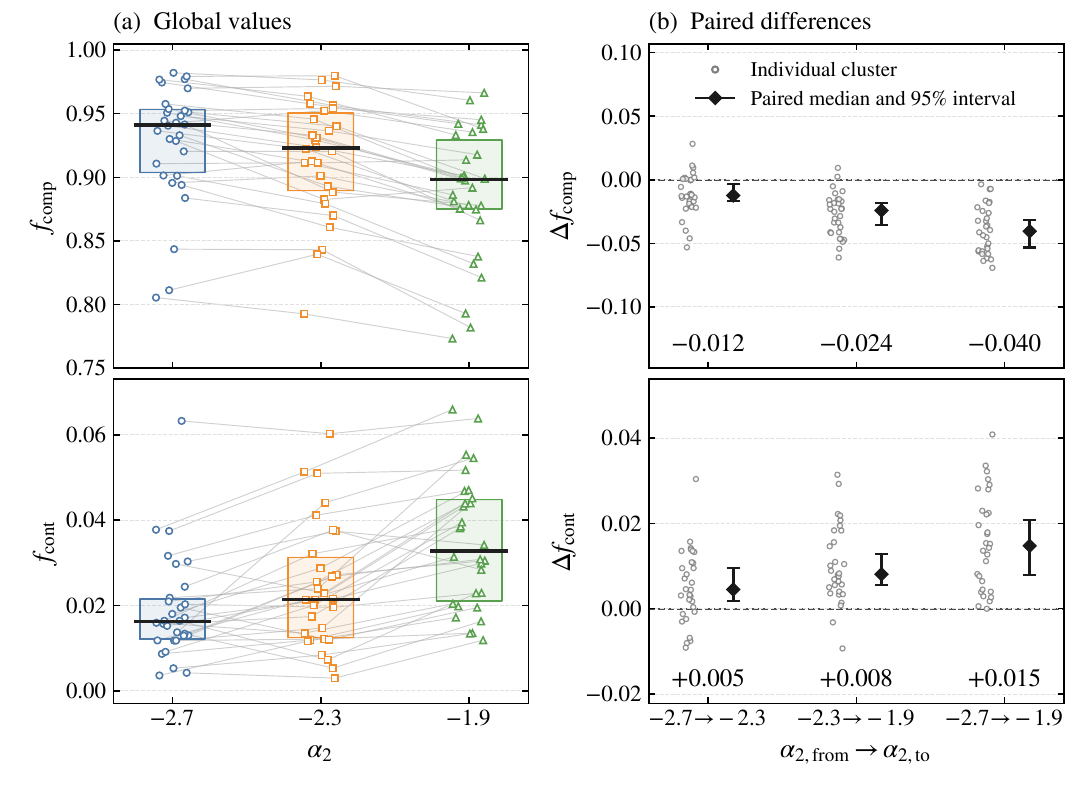}
\centering
\caption{
    Global completeness (top) and contamination (bottom) for the 30 clusters with converged models at all three IMF slopes. (a) Medians over successful realizations for individual clusters. Blue circles, orange squares, and green triangles denote $\alpha_2=-2.7$, $-2.3$, and $-1.9$, respectively; grey lines connect the same cluster. Boxes show the interquartile ranges across clusters, with black bars marking the medians. 
    (b) Within-cluster differences, defined as the value at the second slope minus that at the first. Open grey circles show individual clusters; black diamonds and error bars indicate the median differences and their 95 per cent BCa bootstrap intervals, with the median values given near the bottom of each plot. Dashed lines mark zero.
}
\label{fig:alpha_comparison}
\end{figure*}
Although the selection functions in Fig.~\ref{fig:selection_IMF} do not show a clearly resolved ordering with $\alpha_2$ in individual magnitude bins, such an ordering becomes systematic once the counts are integrated over the full magnitude range. Fig.~\ref{fig:alpha_comparison} shows the resulting global completeness and contamination for the 30 clusters with accepted models at all three IMF slopes. Each point in Fig.~\ref{fig:alpha_comparison}(a) is the median over valid realizations for one cluster at one slope. At fixed $\alpha_2$, the clusters span a wide range, from 0.77 to 0.98 in completeness, with interquartile ranges of 0.05--0.06. Most grey lines linking the same cluster nevertheless decline in completeness and rise in contamination from $\alpha_2=-2.7$ to $-1.9$.

To separate the effect of the IMF from these differences between clusters, we compute, for each cluster, the difference between the values at two slopes, defined as the value at the flatter slope minus that at the steeper one. Fig.~\ref{fig:alpha_comparison}(b) shows these differences and their median across clusters. We estimate a 95 per cent interval on each median from 20\,000 bootstrap resamples of the clusters, using the bias-corrected and accelerated \citep[BCa;][]{efron1987better} method. Between $\alpha_2=-2.7$ and $-1.9$, the median difference is $-0.040$ in completeness, with a BCa interval of $-0.053$ to $-0.031$, and $+0.015$ in contamination, with an interval of $0.008$ to $0.021$. The completeness is lower at $\alpha_2=-1.9$ in all 30 clusters, and the contamination is higher in 29 of them. The two intermediate steps show smaller shifts in the same direction, $-0.012$ and $-0.024$ in completeness, with decreases in 21 and 28 clusters, respectively. 
Thus, even though the models for each cluster are calibrated against the same observed member count, those with flatter IMFs systematically yield lower global completeness and higher global contamination. When the same observed members are corrected for these selection effects, a flatter IMF therefore implies a larger upward correction of the star counts for incompleteness and a larger downward correction for contamination. This is consistent with the signs of the residual normalization bias for wrong-IMF corrections in Fig.~\ref{fig:PDLF_count_shape}(a) and (b).

Although systematic, this IMF-dependent offset is modest compared with the differences between clusters at fixed $\alpha_2$. The median shift of 0.04 in completeness between the two extreme slopes is comparable to the interquartile range across clusters but much smaller than the full range, and the shift of 0.015 in contamination is likewise comparable to its interquartile range of 0.01--0.02. Moreover, the clusters largely retain their relative order across the three slopes, as indicated by the predominantly parallel grey lines in Fig.~\ref{fig:alpha_comparison}(a). We next examine how the differences between clusters relate to their properties.

\begin{figure*}
\includegraphics[width=\textwidth]{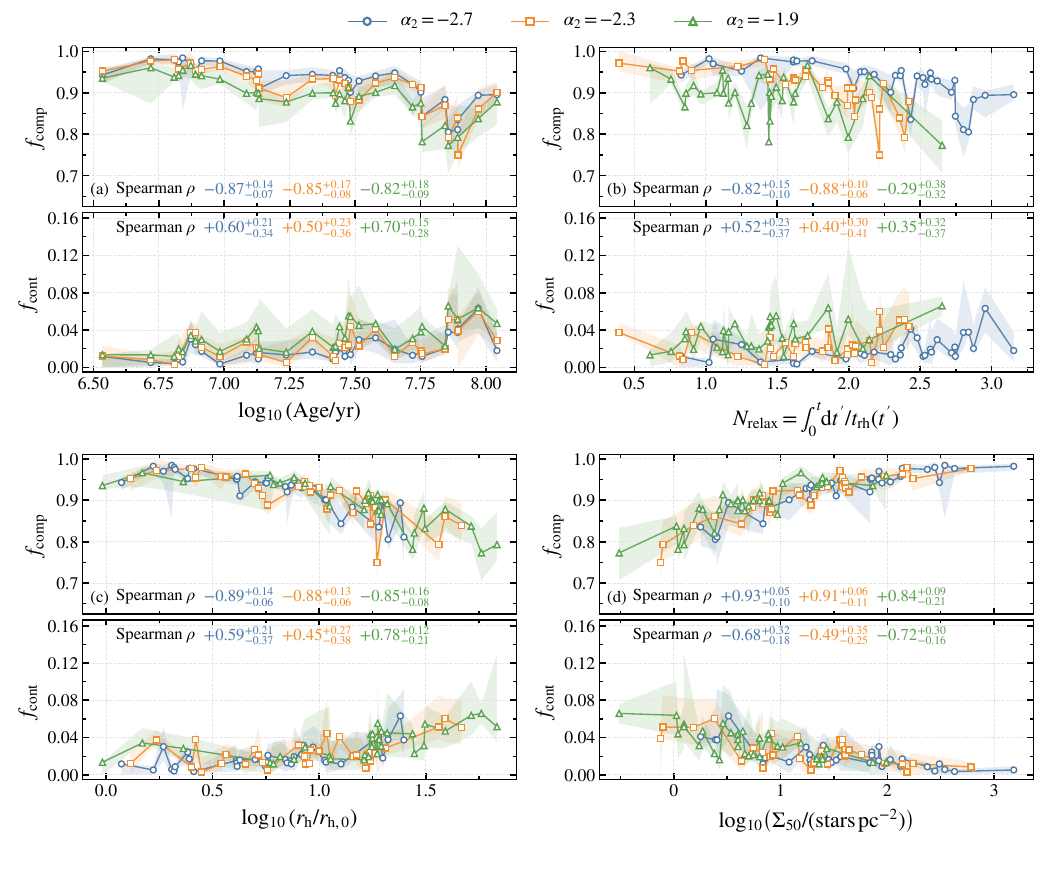}
\centering
\caption{
    Dependence of the recovery completeness, $f_{\rm comp}$ (upper panels), and contamination fraction, $f_{\rm cont}$ (lower panels), on (a) cluster age,(b) the cumulative number of half-mass relaxation times, $N_{\rm relax}=\int_0^t {\rm d}t'/t_{\rm rh}(t')$, (c) the ratio of the half-mass radius at the cluster age to its initial value, $r_{\rm h}/r_{\rm h,0}$, and (d) the projected stellar surface density at the cluster age, $\Sigma_{50}$. Blue circles, orange squares, and green triangles denote $\alpha_2=-2.7$, $-2.3$, and $-1.9$, respectively. Points and shading show the medians and 16th--84th percentile ranges of $f_{\rm comp}$ and $f_{\rm cont}$ across successful realizations of each model. 
    Colour-coded annotations give the Spearman rank correlation coefficients of the model medians, with superscripts and subscripts giving the offsets of the 95 per cent BCa intervals.
}
\label{fig:dependence_properties}
\end{figure*}

Fig.~\ref{fig:dependence_properties}(a) shows that older simulated clusters are recovered less completely and with more contamination. The Spearman coefficients ($\rho$) with age are $-0.87$, $-0.85$, and $-0.82$ for completeness and 0.60, 0.50, and 0.70 for contamination, for $\alpha_2=-2.7$, $-2.3$, and $-1.9$, respectively; all six 95 per cent BCa intervals exclude zero. However, age only sets how long each model has evolved. Clusters of the same age can be at very different stages of dynamical evolution, depending on their mass and size, so age alone does not identify the underlying physical mechanism.

We therefore further consider a dynamical clock, the cumulative number of half-mass relaxation times, $N_{\rm relax}=\int_0^t{\rm d}t'/t_{\rm rh}(t')$, where $t$ is the cluster age and $t_{\rm rh}(t')$ is the half-mass relaxation time \citep{spitzer1987dynamical} of the simulated cluster at time $t'$.\footnote{$t_{\rm rh}=0.138\,N^{1/2}r_{\rm h}^{3/2}/[(G\bar{m})^{1/2}\ln(\gamma N)]$, where $N$, $\bar{m}$, and $r_{\rm h}$ are the number, mean mass, and half-mass radius of all particles of the simulated cluster at time $t'$, and $\ln(\gamma N)$ is the Coulomb logarithm with $\gamma=0.02$, as implemented in {\petar} \citep{wang2020petar}.} For $\alpha_2=-2.7$ and $-2.3$, completeness decreases with $N_{\rm relax}$ about as strongly as with age, with Spearman coefficients of $-0.82$ and $-0.88$. This is consistent with cumulative two-body relaxation contributing to the structural evolution that makes the simulated clusters progressively more difficult to recover. For $\alpha_2=-1.9$, however, the coefficient is only $-0.29$, with an interval of $-0.60$ to $0.10$, and the contamination coefficients are also weaker. One possible reason for the weaker relation at $\alpha_2=-1.9$ is that $N_{\rm relax}$ measures the progress of two-body relaxation, whereas a flatter IMF contains more massive stars, whose stellar evolution mass loss may also contribute to the early expansion of the cluster. This expansion is not driven by two-body relaxation; moreover, the increase in $r_{\rm h}$ lengthens $t_{\rm rh}$, reducing the number of relaxation times accumulated by the same age. For the flatter IMF, $N_{\rm relax}$ may therefore be a less reliable proxy for the cluster structure relevant to membership recovery. This suggests that $N_{\rm relax}$ alone may not consistently explain differences in recovery between clusters across the three IMF assumptions.

Rather than relying on age or a dynamical clock as proxies, we next examine the structural evolution of the simulated clusters directly. Fig.~\ref{fig:dependence_properties}(c) uses the ratio $r_{\rm h}/r_{\rm h,0}$ of the half-mass radius at the cluster age to its initial value. More strongly expanded models show lower completeness for all three IMF slopes, with coefficients of $-0.89$, $-0.88$, and $-0.85$, and higher contamination, with coefficients of 0.45--0.78. Unlike $N_{\rm relax}$, the expansion remains closely related to membership recovery even for $\alpha_2=-1.9$. This suggests that the structural state of the model, as traced by its expansion, is more consistently associated with recovery than the dynamical clock alone.

Finally, Fig.~\ref{fig:dependence_properties}(d) uses the projected stellar surface density $\Sigma_{50}=0.5N_{\mathrm{cl}}^{\mathrm{sim}}/(\upi R_{50}^2)$, where $R_{50}$ is the projected radius enclosing half of the simulated cluster members. Whereas $r_{\rm h}/r_{\rm h,0}$ measures how much a model has expanded relative to its initial state, $\Sigma_{50}$ measures how compact it is at the cluster age. Completeness increases with $\Sigma_{50}$, with coefficients of 0.93, 0.91, and 0.84, and contamination decreases, with coefficients of $-0.68$, $-0.49$, and $-0.72$. The absolute coefficients are comparable to or slightly larger than those for age and expansion, although their confidence intervals overlap. These trends are consistent with the density-based nature of {\hdbscan}: a more diffuse stellar distribution provides less density contrast against the field, making cluster members more difficult to separate from surrounding sources. $\Sigma_{50}$ may therefore serve as a link between the evolution traced in Fig.~\ref{fig:dependence_properties}(a)--(c) and the recovery performance.

Taken together, the four panels of Fig.~\ref{fig:dependence_properties} show that recovery is strongly related to age, expansion, and surface density for all three IMF slopes, but not uniformly to $N_{\rm relax}$. These relations are consistent with a picture in which models expand and become less dense as they evolve, making their members harder to separate from field stars. 
Because the four quantities are correlated with one another, however, the present analysis cannot isolate the effect of any single one.

\section{Discussion}
\label{sec:discussion}

\subsection{Effect of the assumed IMF}
\label{sec:discussion_imf}
 \begin{figure*}
\includegraphics[width=0.9\textwidth]{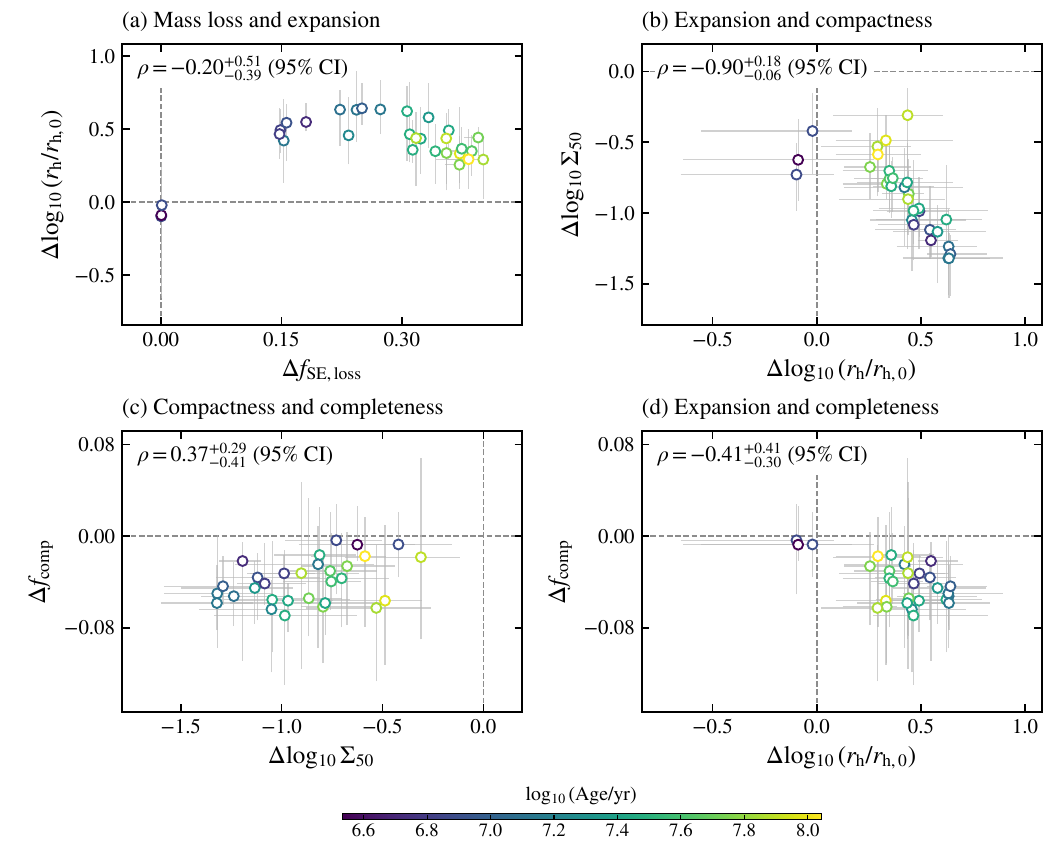}
\centering
\caption{
    Changes in stellar evolution mass loss, structure, and global completeness from $\alpha_2=-2.7$ to $-1.9$ for the 30 clusters with converged models at both IMF slopes. All axes show $\Delta X=X(\alpha_2=-1.9)-X(\alpha_2=-2.7)$, so negative values indicate lower values for the flatter IMF. The panels compare (a) stellar evolution mass loss fraction, $f_{\rm SE,loss}$, with relative expansion, $r_{\rm h}/r_{\rm h,0}$; (b) relative expansion with projected stellar density, $\Sigma_{50}$; (c) projected density with completeness; and (d) relative expansion with completeness. All quantities are medians over the valid realizations of each model, and the changes are differences between these medians, using their logarithms for the structural quantities. Open circles represent individual clusters, coloured by age. Grey bars show the 16th--84th percentile ranges of differences from 20\,000 random pairs of valid realizations, one from each model, indicating realization scatter rather than uncertainties on the medians. In each panel, $\rho$ is the Spearman rank correlation coefficient of the open circles, with superscripts and subscripts giving the offsets of its 95 per cent BCa interval. Dashed lines mark zero change.
}
\label{fig:IMF_effect}
\end{figure*}

We now examine whether the systematic completeness offset between the different IMF assumptions is accompanied by changes in the evolution and structure of the simulated clusters. For each cluster, Fig.~\ref{fig:IMF_effect} shows the changes from $\alpha_2=-2.7$ to $-1.9$ in stellar evolution mass loss, structure, and global completeness. Because both models of a cluster share its age, orbit, field population, and observed member count, these changes remove the offsets between clusters seen in Fig.~\ref{fig:dependence_properties}, although they still include the effect of the different initial masses calibrated for the two IMF slopes.

We quantify the stellar evolution mass loss by $f_{\rm SE,loss}=1-M_{\rm all}(t)/M_{\rm cl,0}$, the cumulative fraction of the initial cluster mass lost by the cluster age $t$ through stellar and binary evolution, including mass ejected in mergers; $M_{\rm all}(t)$ is the total mass of all particles in the simulation at $t$, including remnants and escapers, so that stars escaping from the cluster are not counted as mass loss. As shown in Fig.~\ref{fig:IMF_effect}(a), in 27 of the 30 clusters the flatter-IMF model has a larger $f_{\rm SE,loss}$, with $\Delta f_{\rm SE,loss}\approx0.15$--$0.4$, and expands more relative to its initial half-mass radius, with $\Delta\log(r_{\rm h}/r_{\rm h,0})\approx0.25$--$0.65$. A flatter IMF places a larger fraction of the cluster mass in massive stars, leading to stronger stellar evolution mass loss and, consequently, greater cluster expansion \citep[e.g.][]{kouwenhoven2014how,haghi2020lifetimes}. However, the amount of additional mass loss does not scale monotonically with the amount of additional expansion, because the resulting dynamical response also depends on how rapidly and where the mass is lost \citep[e.g.][]{hills1980effect,alexander2014prescription}.

Regardless of the detailed origin of the expansion, its effect on the present-day structure is clear. 
Clusters for which the flatter-IMF model expands more strongly also show a systematically larger decrease in $\Sigma_{50}$, as in Fig.~\ref{fig:IMF_effect}(b). The Spearman coefficient is $-0.90_{-0.06}^{+0.18}$, and most clusters lie along a narrow sequence. Thus, the differences in expansion between the two IMF assumptions are reflected in the projected stellar densities of the stars passed to membership recovery.

Fig.~\ref{fig:IMF_effect}(c) shows that larger decreases in $\Sigma_{50}$ tend to accompany larger decreases in completeness, and Fig.~\ref{fig:IMF_effect}(d) shows a similar tendency for models that expand more. 
However, the evidence for these trends is limited. The Spearman coefficients are $0.37$ and $-0.41$, with 95 per cent intervals of $-0.04$ to $0.66$ and $-0.71$ to $0.00$, respectively. 
For $\Sigma_{50}$, this is likely because {\hdbscan} recovery depends on the density contrast between cluster and field stars in the five-dimensional space $(l,b,\varpi,\mu_{\alpha *},\mu_\delta)$, which a projected cluster density alone cannot describe.

Three clusters, ASCC~19, OC~0280, and Alessi--Teutsch~10, have $\Delta f_{\rm SE,loss}$, $\Delta\log(r_{\rm h}/r_{\rm h,0})$, and $\Delta f_{\rm comp}$ all close to zero in Fig.~\ref{fig:IMF_effect}. They are younger than 8~Myr and have the lowest calibrated initial masses in our sample, below $60\ M_\odot$ under both IMF assumptions. Under optimal sampling, these low initial masses imply low maximum stellar masses through equation~(\ref{eq:maximum_stellar_mass}). For the resulting stellar mass distributions, little stellar evolution mass loss has occurred by the cluster ages, even with the flatter IMF. By contrast, other clusters of similar age but higher initial mass already show $\Delta f_{\rm SE,loss}\approx0.15$--$0.18$ and $\Delta\log(r_{\rm h}/r_{\rm h,0})\approx0.5$. This contrast suggests that the changes in mass loss and expansion between the two IMF assumptions depend on cluster mass as well as age.

Overall, Fig.~\ref{fig:IMF_effect} shows that models with $\alpha_2=-1.9$ generally undergo more stellar evolution mass loss and expand more than those with $\alpha_2=-2.7$. Their stronger expansion is closely reflected in lower projected stellar densities, so these models are more diffuse. The systematically lower global completeness at $\alpha_2=-1.9$ is then consistent with more diffuse clusters being harder to recover. 
However, clusters with similar structural changes can still differ considerably in their completeness shifts, suggesting that structure is not the only factor behind the completeness difference between the two slopes.

\subsection{Algorithm dependence and applicability of the framework}
\label{sec:discussion_algorithms}
Our selection functions are specific to {\hdbscan} with the settings described in Section~\ref{sec:membership_calibration}. Open cluster member lists, however, are constructed with a variety of methods, including {\dbscan} \citep{castro-ginard2020hunting,castro-ginard2022hunting}, fastMP \citep{perren2023unified}, py\textsc{upmask} \citep{pera2021pyupmask,qin2023hunting}, {\stargo} \citep{yuan2018stargo,pang20223d}, and other machine-learning approaches \citep{vangroeningen2023machinelearningbased}. Because these methods identify members by different criteria, the resulting lists would differ in their completeness at the faint end and in their levels of field contamination. Even with {\hdbscan}, changing the hyperparameters or probability threshold can alter the recovered sample. Selection functions must therefore be derived for the specific procedure used to construct each member list.

The forward-modelling framework developed in this work extends naturally to these algorithm-specific selection functions. Of the four steps in Fig.~\ref{fig:framework}, only the membership recovery in step~3 depends on the algorithm. To derive the selection functions of a given member list, step~3 should reproduce the procedure used to construct that list, applied to the simulated sample with the same input variables, preprocessing, hyperparameters, and member selection criteria as for the observed data. The observational inputs to steps~1 and 2 and the initial cluster mass $M_{\rm cl,0}$ are then taken from, or recalibrated against, that list. Thus, although the selection functions are specific to one algorithm, the framework from which they are derived can be generalized to others.

Because the origin of each simulated star is known, the framework can also be used to evaluate how well candidate algorithms or settings recover a given cluster, and thus to choose the procedure for re-determining its membership. Given the close relation between {\hdbscan} recovery and cluster structure found in Section~\ref{sec:global}, such evaluations should use simulations of the target cluster rather than a generic mock sample.

\subsection{Caveats}
\label{sec:discussion_caveats}
The selection functions derived here depend on several model choices. We vary only the IMF slope above $0.5\,M_\odot$, keeping the slope below this mass fixed and assigning stellar masses by optimal sampling, so that the differences between the models of a cluster can be attributed to $\alpha_2$. The slope below $0.5\,M_\odot$, which governs the number of the faintest members, is therefore not tested. Optimal sampling also fixes the maximum stellar mass for a given $M_{\rm cl,0}$ through equation~(\ref{eq:maximum_stellar_mass}). In the three clusters with the lowest $M_{\rm cl,0}$, this mass is too low for the models with $\alpha_2=-2.7$ and $-1.9$ to differ appreciably in stellar evolution mass loss, expansion, or completeness, as shown in Section~\ref{sec:discussion_imf}. Stochastic sampling could add more massive stars to such clusters, making their selection functions more sensitive to $\alpha_2$, and would increase the scatter between realizations.

The initial half mass radii follow the relation of \citet{marks2012inverse}, which describes embedded clusters before gas expulsion, whereas our models start gas-free and in virial equilibrium, without primordial binaries or mass segregation. They may therefore remain more compact than the observed clusters \citep{zheng2026binary}, many of which are expanding \citep{dellacroce2024young}. As completeness increases with $\Sigma_{50}$ in Fig.~\ref{fig:dependence_properties}(d), such models would tend to overestimate the completeness, and the offset between $\alpha_2=-2.7$ and $-1.9$ found in Section~\ref{sec:discussion_imf} may also depend on the adopted initial structure. Gas expulsion and primordial binaries can be added to the $N$-body models without changing the rest of the framework.

As noted in Section~\ref{sec:membership_recovery}, our membership recovery follows the search used to construct the observed member lists only approximately, and $P_{\rm best}$ is chosen using simulated truth, which is unavailable for observed data. The selection functions therefore describe recovery with an optimally chosen threshold rather than with the criteria of the observed lists. Reproducing the member search of the observed catalogue in full, with the same input variables, preprocessing, hyperparameters, and member selection criteria, would give selection functions that apply directly to the observed lists.

Finally, the calibration matches only $N_{\mathrm{sel}}^{\mathrm{obs}}$. This fixes $M_{\rm cl,0}$ for each assumed IMF but cannot distinguish between IMFs or uniquely determine the initial structure. The choice of IMF nevertheless matters: Fig.~\ref{fig:PDLF_count_shape}(c) and (d) show that corrections derived with $\alpha_2=-2.3$ leave residual distortions in the LF shapes of the target models with $\alpha_2=-1.9$ and $-2.7$. When correcting an observed cluster, the selection functions should therefore be inferred jointly with its IMF, rather than taken from a model with an IMF assumed in advance. 
Such joint inference requires observables beyond the member count. Matching the luminosity function and density profile of the selected members would constrain $\alpha_2$ and $r_{\rm h,0}$, respectively; with a larger grid of $N$-body models, the framework could then infer these initial conditions together with $M_{\rm cl,0}$ and the corresponding selection functions.

\section{Conclusions}
\label{sec:conclusions}

In this work, we developed an IMF-aware forward-modelling framework to quantify the membership selection functions---namely, completeness and field star contamination---of open clusters. We evolve cluster models with $N$-body simulations \citep{wang2020petar}, convert their outputs into synthetic \textit{Gaia} observations, and add mock field stars \citep{rybizki2020gaia}. We then recover cluster members with {\hdbscan} \citep{campello2013densitybased, hunt2023improvinga} to quantify membership completeness and contamination, examining how these quantities vary with the assumed stellar initial mass function (IMF) and relate to cluster evolution.

Applying our framework to 35 nearby young open clusters with three assumed high-mass ($\geq 0.5 M_\odot$) IMF slopes ($\alpha_2 = -2.7, -2.3, -1.9$), we find the following main results:

\begin{itemize}
    \item \textbf{Magnitude- and cluster-dependent selection:} Membership completeness and contamination are highly magnitude-dependent and vary between individual clusters. Faint members are systematically harder to separate from field stars, confirming that accurate derivations of the present-day mass function require corrections that are specific to both the cluster and the precise membership identification procedure used.
    
    \item \textbf{IMF dependence of selection functions:} The global completeness and contamination depend systematically on the assumed IMF. Models initialized with a flatter IMF ($\alpha_2 = -1.9$) contain a larger fraction of massive stars, leading to stronger stellar evolution mass loss. Consequently, these clusters expand more \citep{chernoff1990evolution, kouwenhoven2014how, haghi2020lifetimes, wang2021impact} and exhibit lower projected stellar surface densities at the present day. This reduced density contrast against the field makes members more difficult to recover, yielding systematically lower global completeness and higher global contamination compared to models with steeper IMFs.
    
    \item \textbf{Impact on the recovered luminosity function:} Applying completeness and contamination corrections derived from a mismatched IMF can largely remove the normalization bias (recovering the total number of stars), but it leaves residual distortions in the shape of the present-day luminosity function.
    
    \item \textbf{Correlation with structural evolution:} Cluster recovery performance is strongly tied to the structural state of the model. Completeness significantly decreases and contamination increases for models that have expended more relative to their initial radii and for those with lower present-day projected stellar surface densities, consistent with the density-based nature of {\hdbscan}. 
\end{itemize}

Overall, our results suggest that membership selection functions cannot be universally assumed or derived from generic mock clusters. Because a cluster's evolution and present-day structure are sensitive to its initial mass distribution, correcting an observed cluster's mass function using selection functions derived from an mismatched IMF would inherently bias the result. 
To accurately determine the IMF of an open cluster, these membership selection effects should ultimately be inferred jointly with the IMF and other initial cluster conditions, by incorporating membership selection into $N$-body inferences of these conditions \citep{baumgardt2023evidence, weis2025dynamical, hobart2026initial, zheng2026binary}.

\section*{Acknowledgements}
The authors appreciate the valuable comments and suggestions from the reviewer.
This work is supported by the National Key Research and Development Program of China No. 2025YFF0511001.
L.W. thanks the support from the National Natural Science Foundation of China through grant 12573041 and 12233013, the GuangDong Basic and Applied Basic Research Foundation (2026A1515012460), the High-level Youth Talent Project (Provincial Financial Allocation) through the grant 2023HYSPT0706, the Fundamental Research Funds for the Central Universities, Sun Yat-sen University (2025QNPY04).
X.P. acknowledges the financial support of the National Natural Science Foundation of China through grants 12573036 and 12233013, and the China Manned Space Program with grant No. CMS-CSST-2025-A08.
H.T. is supported by the National Key R\&D Program of China No. 2025YFF0511002 and 2024YFA1611902 and the China Manned Space Project with grant No. CMS-CSST-2025-A11.
This work has made use of data from the European Space Agency (ESA) mission Gaia (https://www.cosmos.esa.int/gaia), processed by the Gaia Data Processing and Analysis Consortium (DPAC, https://www.cosmos.esa.int/web/gaia/dpac/consortium). Funding for the DPAC has been provided by national institutions, in particular the institutions participating in the Gaia Multilateral Agreement.

\section*{Data Availability}
 
Data are available upon request.



\bibliographystyle{mnras}
{\interlinepenalty=10000\relax
\bibliography{refs} 
}




\appendix

\section{Cluster identifications and adopted properties}
\label{app:observation_properties}

Table~\ref{tab:observation_properties} lists the identifications and properties of the 35 clusters in the observational sample described in Section~\ref{sec:observation_sample}. All numerical values are taken from the catalogue of \citet{hunt2024improving}; the names used by \citet{pang20223d} are included to document the sample selection. We match these names against the \texttt{AllNames} column of the \citet{hunt2024improving} catalogue. Where multiple candidate entries are found, we adopt the entry with the smallest angular separation, absolute distance difference, and proper motion difference from the corresponding object in \citet{pang20223d}.

Ages from \citet{hunt2024improving} are expressed in linear units by converting the catalogue columns \texttt{logAge16}, \texttt{logAge50}, and \texttt{logAge84} using $t_q/{\rm Myr}=10^{\ell_q-6}$, where $\ell_q$ denotes the corresponding logarithmic age and $q\in\{16,50,84\}$. They are reported as ${t_{50}}^{+(t_{84}-t_{50})}_{-(t_{50}-t_{16})}$. The simulations use $t_{50}$ rounded down to the nearest Myr.

The member count $N_{\mathrm{cat}}^{\mathrm{obs}}$ is taken from column \texttt{N} of the \citet{hunt2024improving} catalogue, before the magnitude and observational footprint cuts described in Section~\ref{sec:observation_sample}. Cluster masses are taken directly from the \texttt{MassTot} column, whose values already include corrections for completeness and unresolved binaries. The quoted mass uncertainties are taken from \texttt{e\_MassTot}. These are present-day total masses, rather than initial cluster masses or masses enclosed within the Jacobi radius.

\input{Parameters_Hunt_Pang_comparison.tex}

\section{GeDR3mock query for field stars}
\label{app:mock_field_query}
Table~\ref{tab:gedr3mock_query} lists the conditions used to retrieve the mock field stars of each cluster from \texttt{gedr3mock.main} through the TAP service of the GAVO Data Center (\url{https://dc.g-vo.org/tap}).

\begin{table}
\centering
\caption{Conditions of the GeDR3mock query for each cluster. The search limits are centred on the footprint centre (Section~\ref{sec:synthetic_gaia}); the subscript d denotes the download limits defined in Section~\ref{sec:mock_field}, and $i\in\{\alpha *,\delta\}$. The magnitude limit is fainter than the adopted $4\leq G\leq19$~mag cut, which is applied after retrieval together with the footprint cuts. The stellar-cluster component of the catalogue is identified by \texttt{popid}~$=11$.}
\label{tab:gedr3mock_query}
\begin{tabular}{ll}
\hline
Quantity & Condition \\
\hline
\multicolumn{2}{l}{\textit{Search limits}} \\[2pt]
Sky position & $\theta\leq\theta_{\rm d}$ \\
Parallax & $|\varpi-\varpi_0|\leq\Delta\varpi_{\rm d}$ \\
Proper motion & $\sum_i\bigl[(\mu_i-\mu_{i,0})/\Delta\mu_{i,{\rm d}}\bigr]^2<4$ \\
Magnitude & $\texttt{phot\_g\_mean\_mag}\leq21$~mag \\[4pt]
\multicolumn{2}{l}{\textit{Quality criteria}} \\[2pt]
Visibility periods & $\texttt{visibility\_periods\_used}>5$ \\
Population & \texttt{popid}~$\neq11$ \\
$G$ flux & $\texttt{phot\_g\_mean\_flux\_over\_error}>10$ \\
$G_{\rm BP}$ flux & $\texttt{phot\_bp\_mean\_flux\_over\_error}>10$ \\
$G_{\rm RP}$ flux & $\texttt{phot\_rp\_mean\_flux\_over\_error}>10$ \\
Parallax & $\texttt{parallax\_over\_error}>10$ \\
\hline
\end{tabular}
\end{table}

\section{HDBSCAN hyperparameter settings}
\label{app:hdbscan_hyperparameters}

\subsection{Fixed hyperparameters}
\label{app:hdbscan_fix}
By default, we adopt $m_{\rm clSize}=10$, $m_{\rm Pts}=5$, and excess of mass (\texttt{eom}) cluster selection for all models. Holding these settings fixed keeps the recovery procedure consistent across mass-calibration trials and limits repeated tuning.

The neighbourhood parameter $m_{\rm Pts}$ sets the scale for local density estimation. Larger values make clustering more conservative, generally assigning more sources to noise and restricting detections to denser regions\footnote{\url{https://hdbscan.readthedocs.io/en/latest/parameter_selection.html}}. We choose $m_{\rm Pts}=5$ as a relatively permissive setting; this value has also been used for open cluster membership identification by \citet{alfonso2024exploring}. The minimum cluster size $m_{\rm clSize}=10$ allows groups with as few as ten sources to remain eligible for recovery, as also adopted by \citet{qin2023hunting}.

The \texttt{eom} method selects structures by their integrated stability in the density hierarchy, while \texttt{leaf} selection produces finer subdivisions. Our aim is to recover the target cluster as a whole, motivating the use of \texttt{eom}. These choices define the recovery procedure for which the selection functions are measured. Each trial must satisfy the recovery criteria in Section~\ref{sec:membership_recovery}; if the mass calibration does not converge with these settings, we perform the supplementary hyperparameter search below.

\subsection{Free hyperparameter search}
\label{app:hdbscan_search}
The supplementary search covers 28 combinations of $m_{\rm clSize}\in\{10,15,20,30,40,50,80\}$, $m_{\rm Pts}\in\{5,10\}$, and \texttt{eom} or \texttt{leaf} selection. For each fit, the target is the non-noise label containing the most true cluster stars; ties favour the smaller label.

We evaluate completeness and contamination at 500 equally spaced membership-strength thresholds over $[0,1]$, sort by contamination, and integrate completeness against contamination using the trapezoidal rule. Empty selections are assigned zero contamination for ranking only. We retain the fit with the largest area; ties favour \texttt{eom} over \texttt{leaf}, then increasing $m_{\rm clSize}$ and $m_{\rm Pts}$. This completeness--contamination area is not a standard receiver-operating-characteristic area. We then select $P_{\rm best}$ using the trial-wide $f_{\rm comp}-f_{\rm cont}$ criterion in Section~\ref{sec:membership_recovery}, without refitting. The supplementary search therefore uses simulated truth for both target and hyperparameter selection, independently of the observed member count.

\section{Selection functions of the full sample}
\label{app:selection_full}
Fig.~\ref{fig:selection_full} shows the selection functions of the 28 clusters with accepted models that are not included in Fig.~\ref{fig:selection_IMF}; together, the two figures cover all 34 clusters used in the selection function analysis. Accepted models are available for only two IMF slopes for NGC~7429, NGC~2451A, and NGC~1980, and only for $\alpha_2=-2.7$ for IC~2602, for which no comparison between slopes is therefore possible. For IC~2602, only three of the 16 realizations satisfy the recovery criteria in equation~(\ref{eq:recovery_criteria}), so its percentile ranges are poorly sampled.

\begin{figure*}
\includegraphics[width=\textwidth]{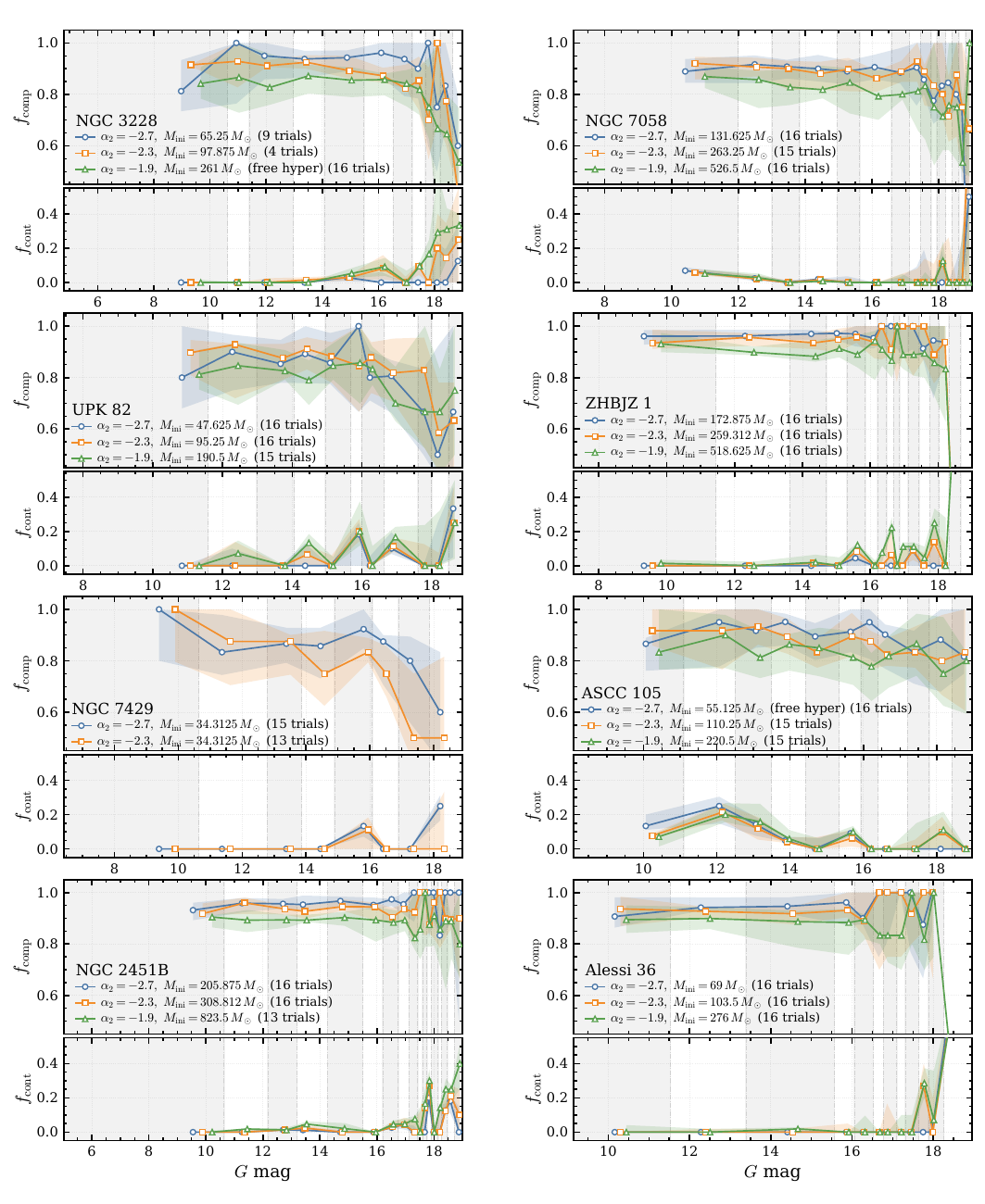}
\caption{
    Selection functions of the 28 clusters with accepted models that are not shown in Fig.~\ref{fig:selection_IMF}. Panels, symbols, curves, shading, and legends are as in Fig.~\ref{fig:selection_IMF}.
}
\label{fig:selection_full}
\end{figure*}

\begin{figure*}
\includegraphics[width=\textwidth]{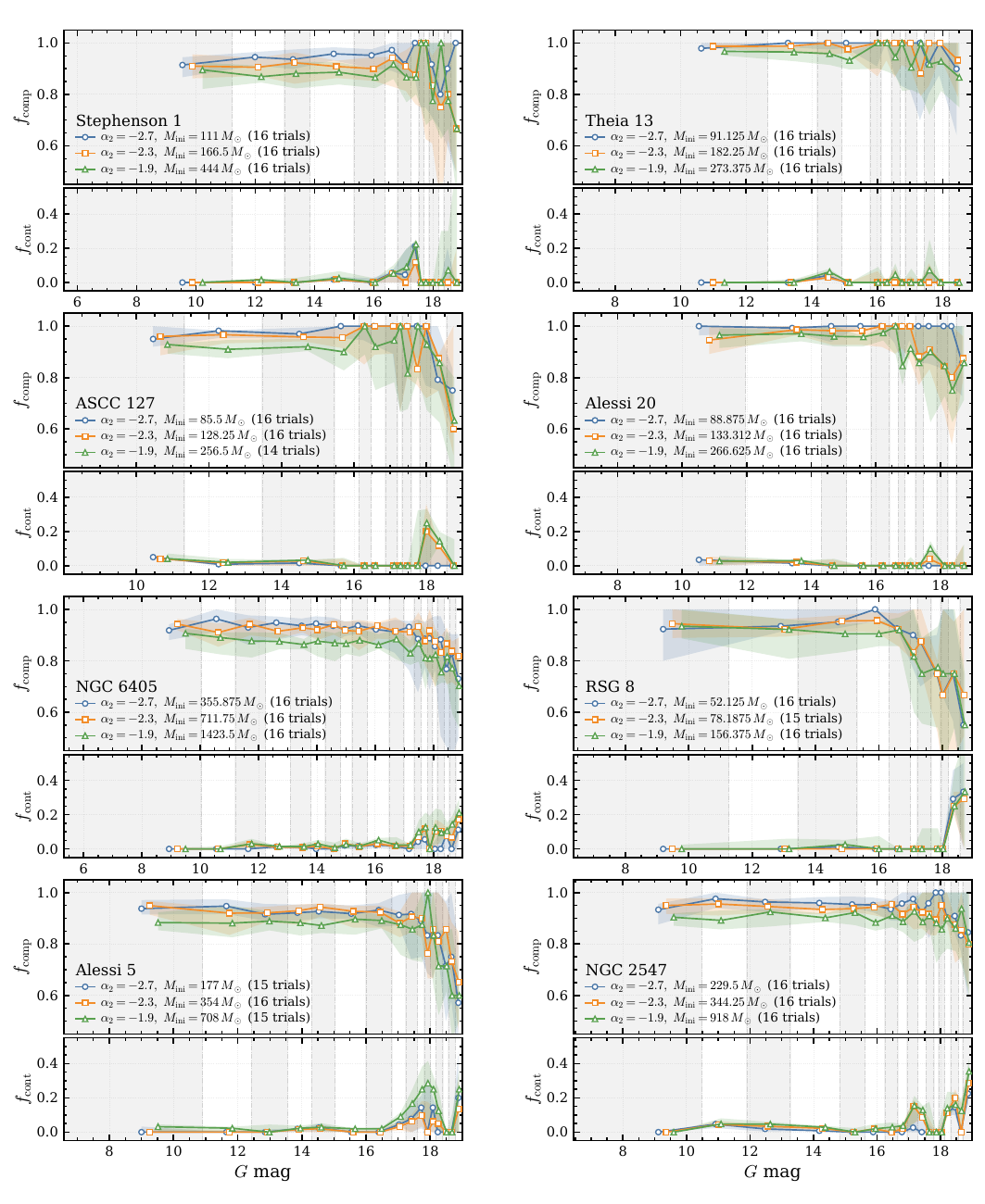}
\contcaption{}
\end{figure*}

\begin{figure*}
\includegraphics[width=\textwidth]{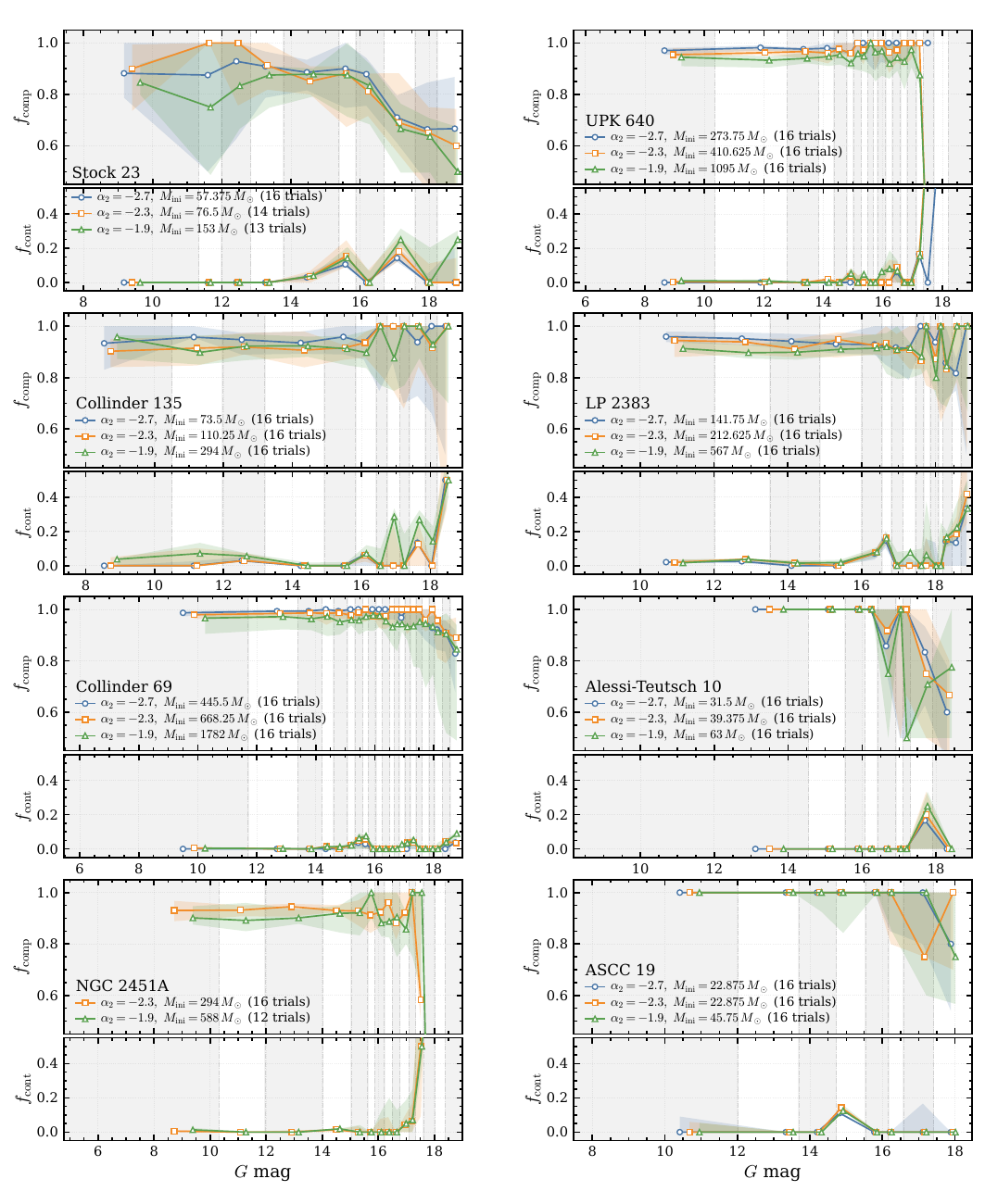}
\contcaption{}
\end{figure*}

\begin{figure*}
\includegraphics[width=\textwidth]{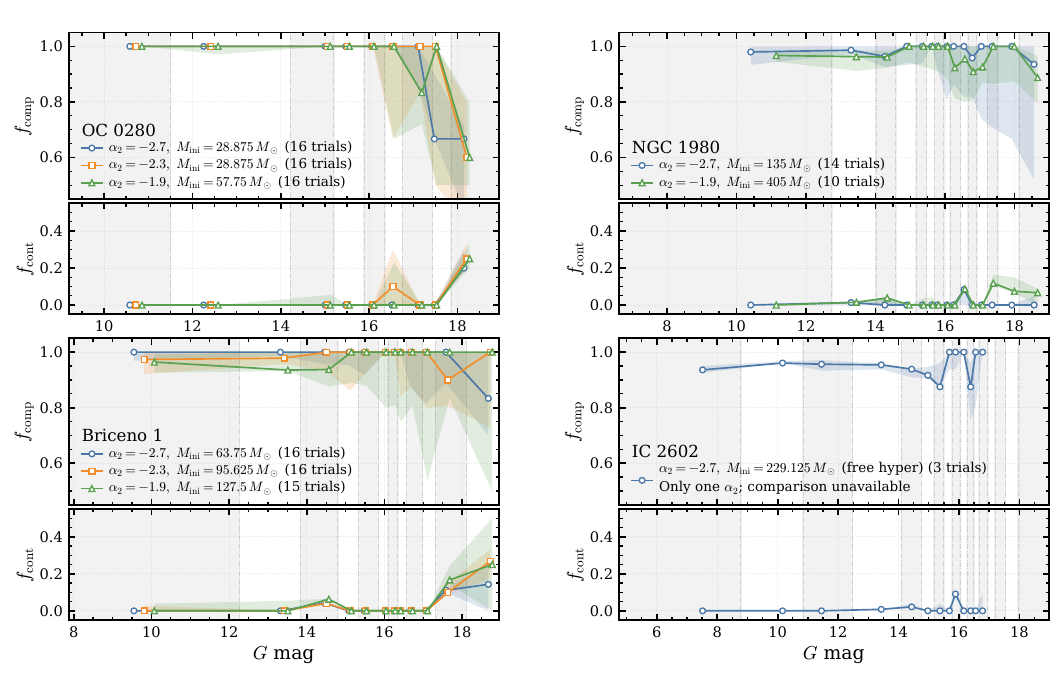}
\contcaption{}
\end{figure*}


\bsp	
\label{lastpage}
\end{document}

%% file: Parameters_Hunt_Pang_comparison.tex
\begin{table*}
\centering
\makeatletter\let\refstepcounter\H@refstepcounter\makeatother
\caption{Identifications and HR24 properties of the 35 clusters in the observational sample. The P22 name column gives the corresponding names used in selecting the sample; all numerical values are from HR24. Ages are shown as medians with the 16th--84th percentile intervals after conversion to linear units. See Appendix~\ref{app:observation_properties} for the matching procedure and property definitions.}
\label{tab:observation_properties}
\begingroup
\footnotesize
\setlength{\tabcolsep}{4pt}
\renewcommand{\arraystretch}{0.98}
\begin{tabular}{llrrrrrrrr}
\hline
HR24 name & P22 name & R.A. & Decl. & $d$ & $\mu_{\alpha*}$ & $\mu_\delta$ & $t$ & $M_{\rm cl}$ & $N_{\mathrm{cat}}^{\mathrm{obs}}$ \\
 & & (deg) & (deg) & (pc) & (mas yr$^{-1}$) & (mas yr$^{-1}$) & (Myr) & ($\mathrm{M}_\odot$) & \\
\hline
ASCC~105 & ASCC~105 & 295.472 & 27.386 & 553.4 & 1.370 & -1.618 & $93.4_{-40.9}^{+73.0}$ & $307.3\pm22.4$ & 173 \\[2pt]
ASCC~127 & ASCC~127 & 347.302 & 65.183 & 374.2 & 7.410 & -1.645 & $12.2_{-5.1}^{+7.1}$ & $168.6\pm20.4$ & 247 \\[2pt]
Briceno~1 & ASCC~16 & 81.083 & 1.696 & 342.1 & 1.384 & -0.016 & $8.2_{-3.8}^{+3.1}$ & $118.5\pm19.3$ & 171 \\[2pt]
ASCC~19 & ASCC~19 & 81.978 & -1.856 & 352.1 & 1.168 & -1.182 & $7.4_{-2.7}^{+3.6}$ & $56.1\pm13.0$ & 61 \\[2pt]
ASCC~58 & ASCC~58 & 153.706 & -55.020 & 471.4 & -13.302 & 2.760 & $57.0_{-26.7}^{+48.1}$ & $266.5\pm32.9$ & 236 \\[2pt]
Alessi~20 & Alessi~20 & 2.643 & 58.757 & 424.3 & 8.117 & -2.427 & $6.7_{-2.8}^{+2.3}$ & $204.7\pm26.6$ & 252 \\[2pt]
Alessi~5 & Alessi~5 & 160.902 & -61.112 & 391.8 & -15.400 & 2.544 & $56.4_{-22.7}^{+40.1}$ & $657.9\pm53.6$ & 571 \\[2pt]
BH~99 & BH~99 & 159.554 & -59.109 & 440.2 & -14.477 & 0.980 & $44.8_{-19.1}^{+42.7}$ & $739.4\pm60.2$ & 643 \\[2pt]
Collinder~135 & Collinder~135 & 109.406 & -36.921 & 295.9 & -10.078 & 6.230 & $29.8_{-13.1}^{+23.9}$ & $128.3\pm25.7$ & 209 \\[2pt]
Collinder~140 & Collinder~140 & 110.802 & -32.012 & 376.2 & -8.088 & 4.756 & $17.3_{-7.7}^{+12.6}$ & $164.2\pm22.8$ & 203 \\[2pt]
Collinder~69 & Collinder~69 & 83.814 & 9.888 & 393.8 & 1.215 & -2.047 & $5.2_{-1.8}^{+1.8}$ & $510.6\pm92.7$ & 1247 \\[2pt]
Theia~13 & Gulliver~6 & 83.132 & -1.753 & 408.8 & 0.104 & -0.222 & $6.4_{-2.4}^{+3.1}$ & $202.3\pm25.9$ & 249 \\[2pt]
IC~2391 & IC~2391 & 130.260 & -53.042 & 150.2 & -24.797 & 23.267 & $27.3_{-12.9}^{+20.2}$ & $203.9\pm27.6$ & 376 \\[2pt]
IC~2602 & IC~2602 & 160.973 & -64.393 & 150.6 & -17.775 & 10.669 & $26.0_{-10.8}^{+21.6}$ & $344.2\pm42.3$ & 638 \\[2pt]
IC~4665 & IC~4665 & 266.532 & 5.635 & 343.3 & -0.881 & -8.532 & $32.7_{-14.4}^{+20.2}$ & $278.9\pm25.2$ & 308 \\[2pt]
LP~2383 & LP~2383 & 95.380 & -16.097 & 365.0 & -5.370 & 5.049 & $37.9_{-19.6}^{+40.9}$ & $322.4\pm33.9$ & 422 \\[2pt]
ZHBJZ~1 & LP~2439 & 103.947 & -5.791 & 283.7 & -7.318 & -2.445 & $13.5_{-6.9}^{+8.8}$ & $275.0\pm37.2$ & 490 \\[2pt]
UPK~640 & LP~2442 & 249.692 & -39.507 & 173.9 & -11.882 & -21.391 & $9.6_{-4.4}^{+4.8}$ & $484.1\pm44.0$ & 764 \\[2pt]
NGC~1980 & NGC~1980 & 83.801 & -5.921 & 378.3 & 1.265 & 0.472 & $6.9_{-2.6}^{+2.9}$ & $245.9\pm35.5$ & 364 \\[2pt]
NGC~2232 & NGC~2232 & 96.978 & -4.772 & 318.0 & -4.717 & -1.770 & $13.3_{-5.6}^{+9.2}$ & $185.8\pm25.5$ & 286 \\[2pt]
NGC~2451A & NGC~2451A & 115.971 & -38.241 & 190.1 & -20.987 & 15.327 & $26.1_{-12.9}^{+22.7}$ & $220.8\pm30.4$ & 407 \\[2pt]
NGC~2451B & NGC~2451B & 116.176 & -37.942 & 367.5 & -9.400 & 4.744 & $27.6_{-10.7}^{+19.5}$ & $494.1\pm52.0$ & 615 \\[2pt]
NGC~2547 & NGC~2547 & 122.533 & -49.195 & 382.4 & -8.568 & 4.331 & $21.7_{-9.2}^{+19.5}$ & $530.4\pm54.3$ & 675 \\[2pt]
NGC~3228 & NGC~3228 & 155.278 & -51.845 & 477.1 & -14.856 & -0.595 & $30.3_{-12.1}^{+25.8}$ & $250.8\pm23.2$ & 200 \\[2pt]
NGC~6405 & NGC~6405 & 265.067 & -32.228 & 451.9 & -1.368 & -5.847 & $52.4_{-22.3}^{+38.5}$ & $1531.7\pm173.4$ & 974 \\[2pt]
NGC~7058 & NGC~7058 & 320.409 & 50.830 & 362.8 & 7.429 & 2.805 & $69.7_{-35.3}^{+52.9}$ & $334.1\pm43.2$ & 424 \\[2pt]
NGC~7429$^{\rm m}$ & RSG~7 & 344.090 & 59.912 & 417.4 & 4.771 & -2.245 & $78.1_{-32.0}^{+58.5}$ & $71.8\pm14.8$ & 67 \\[2pt]
RSG~8 & RSG~8 & 345.453 & 59.509 & 481.4 & 5.664 & -1.716 & $13.6_{-5.5}^{+8.8}$ & $175.1\pm19.1$ & 151 \\[2pt]
Roslund~5 & Roslund~5 & 302.617 & 33.741 & 538.9 & 2.052 & -1.236 & $110.1_{-41.8}^{+95.0}$ & $489.1\pm43.1$ & 360 \\[2pt]
Stephenson~1 & Stephenson~1 & 283.578 & 36.881 & 354.7 & 1.121 & -3.018 & $26.6_{-12.1}^{+23.5}$ & $267.7\pm27.2$ & 322 \\[2pt]
Stock~23 & Stock~23 & 49.184 & 60.071 & 598.2 & -4.324 & -0.993 & $71.8_{-30.9}^{+46.7}$ & $203.3\pm37.4$ & 106 \\[2pt]
Alessi-Teutsch~10 & UBC~19 & 56.331 & 29.742 & 392.6 & 2.671 & -5.238 & $3.4_{-0.9}^{+0.5}$ & $46.5\pm9.5$ & 85 \\[2pt]
OC~0280$^{\rm m}$ & UBC~31 & 60.521 & 32.178 & 368.0 & 3.628 & -5.259 & $7.8_{-3.3}^{+3.6}$ & $52.4\pm12.5$ & 78 \\[2pt]
Alessi~36 & UBC~7 & 106.565 & -37.596 & 275.3 & -9.669 & 7.046 & $28.9_{-14.3}^{+31.9}$ & $105.8\pm16.8$ & 198 \\[2pt]
UPK~82 & UPK~82 & 298.464 & 26.202 & 537.9 & 2.278 & -2.169 & $77.9_{-35.1}^{+72.3}$ & $194.7\pm26.5$ & 142 \\[2pt]
\hline
\end{tabular}
\endgroup
\medskip
\parbox{\textwidth}{\footnotesize HR24: \citet{hunt2024improving}; P22: \citet{pang20223d}. Masses and their uncertainties are taken directly from \texttt{MassTot} and \texttt{e\_MassTot}, respectively; the masses already include corrections for completeness and unresolved binaries. $N_{\mathrm{cat}}^{\mathrm{obs}}$ is the HR24 catalogue member count (column \texttt{N}) before the magnitude and observational footprint cuts applied in this work. Superscript m denotes an HR24 moving-group classification.}
\end{table*}